\documentclass{aa}  

\usepackage{graphicx}
\usepackage{txfonts}
\usepackage[colorlinks=true, citecolor=blue]{hyperref}
\newcommand{\Hmol}{\mbox{H$_{\rm 2}$}}

\newcommand{\Msun}{M$_{\odot}$}
\newcommand{\Lsun}{L$_{\odot}$}

\newcommand{\mum}{$\mu$m}

\begin{document} 
   
  \title{Radiative and mechanical energies in galaxies}
  \subtitle{I. Contributions of molecular shocks and PDRs in 3C~326~N}
  
  \author{J. A. Villa-V\'elez
          \inst{1}\fnmsep\thanks{E-mail: \url{jorge.villa@phys.ens.fr}}
          \and
          B. Godard\inst{2,1}\and
          P. Guillard\inst{3}\and
          G. Pineau des Forêts\inst{4,2} }
          
   \institute{Laboratoire de Physique de l’École Normale Supérieure, ENS, Université PSL, CNRS, Sorbonne Université, Université Paris Cité, F-75005 Paris, France
              \and
              Observatoire de Paris, PSL University, Sorbonne Université, LERMA, 75014 Paris, France
              \and
              Sorbonne Université, CNRS, UMR 7095, Institut d’Astrophysique de Paris, 98bis bd Arago, 75014 Paris, France
              \and
              Université Paris-Saclay, CNRS, Institut d'Astrophysique Spatiale, 91405 Orsay, France
             }

   \date{Received \today; accepted \today}


    \abstract {Atomic and molecular lines emitted from galaxies are fundamental tracers of the medium responsible for the emission and contain valuable information regarding the energy budget and the strength of the different feedback mechanisms.} {The goal of this work is to provide a new framework for the interpretation of atomic and molecular lines originating from extragalactic sources and a robust method to deduce the mechanical and radiative energy budget from a set of observations.} {Atomic and molecular lines detected in a given object are assumed to result from the combination of distributions of shocks and photo-dissociation regions (PDR) within the observational beam. The emission of individual structures is computed using the Paris-Durham shock code and the Meudon PDR code over a wide range of parameters. The total emission is then calculated assuming probability distribution functions for shocks and PDRs. A distance between the observational dataset and the model is finally defined based on the ratios of the observed and predicted intensities.} {As a test case scenario, we consider the radio galaxy 3C~326~N. The dataset is composed of 12 rotational and ro-vibrational lines of \Hmol\ and the fine structure lines of C$^+$ and O. Our interpretative framework shows that both shocks and PDRs are required to explain the line fluxes. Surprisingly, viable solutions are obtained at low density only ($\rm n_H < 100~cm^{-3}$), indicating that most of the emission originates from diffuse interstellar matter. The optimal solution, obtained for $\rm n_H = 10~cm^{-3}$, corresponds to a distribution of low-velocity shocks (between $5$ and $20$~km~s$^{-1}$) propagating in PDR environments illuminated by a UV radiation field $10$ times larger than that in the solar neighborhood. This solution implies that at least 4\% of the total mass carried by the PDRs is shocked. The \Hmol\ 0-0 S(0)~$28~\mu$m, [CII]~$158~\mu$m, and [OI]~$63~\mu$m lines originate from the PDR components, while all the other \Hmol\ lines are mostly emitted by shocks. The total solid angles sustained by PDRs and shocks imply that the radiative and mechanical energies reprocessed by these structures are $\mathcal{L}_{\rm UV} = 6.3\times10^9~{\rm L_\odot}$, and $\mathcal{L}_{\rm K} = 3.9\times10^8~{\rm L_\odot}$, respectively, in remarkable agreement with the values of the infrared luminosity deduced from the fit of the spectral energy distribution of 3C~326~N, and consistent with a small fraction of the AGN jet kinetic power dissipated in the interstellar medium ($\approx 1$\%).} {This work shows that the radiative and mechanical energy budget of galaxies can be derived from the sole observations of atomic and molecular lines.  It reveals the unexpected importance of the diffuse medium for 3C~326~N, in contrast to previous studies. A last minute comparison of the model to new JWST data obtained in 3C~326~N show a striking agreement and demonstrate the ability of the model to make accurate predictions. This framework therefore opens new prospects for the prediction and interpretation of extragalactic observations, in particular in the context of JWST observations.}

   \keywords{shock waves --
            ISM: kinematics and dynamics --
            ISM: photon-dominated region (PDR) --
            ISM: molecules --
            Galaxies: ISM --
            Galaxies: jets
               }

   \titlerunning{Radiative and mechanical energy budgets in galaxies}

   \maketitle
%

\section{Introduction}\label{sec:intro}

Galaxies are dynamic systems where diverse energy sources and sinks shape their evolution over cosmic time. The intricate interplay between dynamics, energy dissipation, and star formation constitutes the essence of galaxy formation and evolution \citep[e.g.][]{, Pfenniger2010}. For instance, in the hypothetical absence of radiative losses, the kinetic energy released by gravitational accretion onto halos would keep the gas unbound and galaxies would not form at all \citep{Benson2010}. It is therefore of paramount importance to understand and quantify how and where radiative and mechanical energies are reprocessed and dissipated in the multi-phase gas of galaxies.

The reprocessing of radiative and mechanical energy in the interstellar medium (ISM) regulates the distribution of gas phases \citep[e.g.,][]{Vazquez-Semadeni2009}, the formation of molecular clouds and stars \citep{Klessen2010}, and the feedback mechanisms \citep{Cielo2018, Ostriker2022}. 
Mechanical processes encompass the momentum input from stellar winds \citep{Krumholz2014}, supernovae \citep{Fierlinger2016}, and active galactic nuclei (AGN) \citep{Ciotti2010}. Simultaneously, radiative processes involve cooling through thermal and non-thermal emissions across the electromagnetic spectrum, from radio waves to gamma-rays. The dissipation of radiative and mechanical energy in the ISM is intricately linked to the gas heating and cooling mechanisms, governed by the microphysics. The thermodynamic and chemical states of the gas significantly influence ISM observables, including ionic, atomic, and molecular line emissions. Those gas properties depend on factors such as density, metal and dust abundance, and the strengths of heating/ionization sources like the ultra-violet (UV) radiation field and cosmic rays.

Efforts to model line and continuum emission in galaxies provide a means to constrain dominant gas excitation mechanisms and quantify the reprocessed energy required to reproduce observations. Nonetheless, the complexity of the ISM poses significant challenges in solving radiative transfer within galaxies \citep{Kim2023}. As a result, thermal equilibrium and energy balance arguments are frequently employed to simplify calculations. For instance, contemporary codes used to replicate the panchromatic spectral energy distributions (SEDs) of galaxies often assume that stellar UV and near-infrared (NIR) light is absorbed by interstellar dust and subsequently re-emitted in the far-infrared (FIR) domain \citep{daCunha2008, Carnall2018, Boquien19}, primarily in Photo-Dissociation Regions (PDRs) situated at the interfaces between molecular clouds and diffuse regions of atomic and ionized gas. This assumption is commonly employed to constrain the amount of reprocessed radiative energy from FIR observations and deduce the properties of infrared (IR) galaxies \citep[e.g.,][]{Casey2014}.

The physical modeling of spectral line cooling from galaxies, arising from the de-excitation of ions, atoms, and molecules, typically focuses on specific chemical states of the gas (ionized, atomic, or molecular), ISM phases (particular temperature or density ranges), or energy types (electromagnetic radiation, cosmic rays, or mechanical dissipation). For instance, well-established photo-dissociation codes like CLOUDY \citep{Ferland2017} or the Meudon PDR code \citep{LePetit2006} are commonly employed to model the reprocessing of radiative energy, thereby constraining excitation mechanisms and physical conditions of X-ray or UV-irradiated gas in galaxies \citep[e.g.,][]{Polles2021}. Conversely, the reprocessing of mechanical energy in ionized/atomic or molecular media has been incorporated into codes such as MAPPINGS \citep{Allen2008} or the Paris-Durham shock code \citep{FlowerPdF03, Godard19}. The majority of investigations employing the aforementioned models to characterize molecular line emissions from galaxies conclude that both radiative and mechanical processes contribute substantively to the observed gas excitation. Illustratively, the Spectral Line Energy Distribution (SLED) of the carbon monoxide (CO) molecule is frequently modeled through a mix of Photo-Dissociation Regions (PDRs), X-ray Dominated Regions (XDRs) and shocks \citep{Meijerink13, Mingozzi18, Esposito22}.

This paper addresses the reprocessing of energy in the molecular phase of the ISM. We present a method to estimate the relative contributions of shocks and PDRs to the observed line emission from galaxies, with a focus on \Hmol\ lines. Given that \Hmol\ can form on interstellar dust at elevated temperatures \citep[$\approx$~50 to 250~K,][]{Grieco23} and that the \Hmol\ rotational levels are separated by a few hundreds of Kelvin, it emerges as a significant coolant for warm gas and, consequently, a key participant in the reprocessing of radiation within PDRs \citep{Habart2005, Shaw2009} and the dissipation of kinetic energy within molecular shocks \citep{Flower1999, Kristensen23}. A previous study by \cite{Lesaffre2013} used probability distribution functions (PDFs) of shock velocities to model the rotational and ro-vibrational lines of \Hmol\ in molecular gas heated by shock waves. These emission lines serve as indicators of mechanical energy dissipation \citep{Lesaffre2020}, particularly within galaxies where the total power of IR \Hmol\ line emissions exceeds what can be accounted for by the exclusive reprocessing of the available UV radiative luminosity \citep{Guillard2009, Ogle10, Nesvadba10, Herrera12, Guillard15, Guillard15a}.

This paper presents an extension of the framework presented in \cite{Lehmann22}, where we now combine distributions of shocks and PDRs models. The state-of-art codes used to model the line emission are presented in Sect.~\ref{sec:modelling}. The toy model we use to describe the line fluxes from an ensemble of shocks and PDRs is described in Sect.~\ref{sec:distribution-shocks}, as well as the methodology to fit a suite of emission lines with grids of models. In Sect. \ref{sec:results}, we apply this modeling to the 3C~326~N radio-galaxy, constraining UV- and shock-processed gas mass and energy budgets, and finally, we provide predictions of the suite of ro-vibrational \Hmol\ emission lines that will be observed with the JWST. Throughout this work, we assume a modern flat cosmology \citep{Planck2020} with $\Omega_{\rm m}$ = 0.315, $\Omega_\Lambda$ = 0.685, $\Omega_{\rm b}$ = 0.0493, and H$_0$ = 67.4 km s$^{-1}$ Mpc$^{-1}$.

\section{Shock and PDR models}\label{sec:modelling}

Throughout this paper, two publicly available codes are used to model PDR and shock emissions: the Meudon PDR code\footnote{Meudon PDR code: \url{https://ism.obspm.fr/pdr.html}} \citep{LePetit2006} and the Paris-Durham shock code\footnote{Paris-Durham code: \url{https://ism.obspm.fr/shock.html}} \citep{FlowerPdF03, Godard19}.

\subsection{Grids of PDR models}\label{sec:PDR_models}

The Meudon PDR code is a model designed to study the thermal and chemical structures of photo-dissociation regions with a constant proton density, $\rm n_H$, or a constant thermal pressure. The geometry, the main assumptions, and the physical ingredients included in the code are described in \cite{LePetit2006}. In a nutshell, the model considers a 1D plane-parallel and static slab of gas and dust illuminated on both sides by semi-isotropic radiation fields, each of which is set to the standard interstellar radiation field (ISRF, \citealt{Mathis83}) and scaled with a parameter $\rm G_0$. The size of the cloud is set by its total extinction $\rm A_{v~max}$. Within this geometry, the model solves the radiative transfer, taking into account the detailed absorption and emission processes induced by dust and gas compounds, and the thermal and chemical states of matter at equilibrium, through the inclusion of an extensive chemical network.

In addition to the chemical composition of the gas, the model also solves the excitation of atomic and molecular species including the populations of 250 rovibrational levels of H$_2$. In particular, the excitation of the fine structure levels of C$^+$ and O is computed including excitation by collisions with H, H$_2$, and electrons (for C$^+$) and with H, H$^+$, H$_2$, He, and electrons (for O). The populations of the ro-vibrational levels of H$_2$ are computed taking into account excitation by collisions with H, He, H$_2$, and H$^+$, the probability of exciting H$_2$ at formation on grain surfaces, and the excitation by radiative pumping of the electronic lines of molecular hydrogen followed by fluorescence. The emerging line fluxes of all these species are finally calculated by integrating the emissivities over the entire slab up to $\rm A_{v~max}$.

\begin{figure*}
    \centering
     \includegraphics[width=0.99\textwidth,clip]{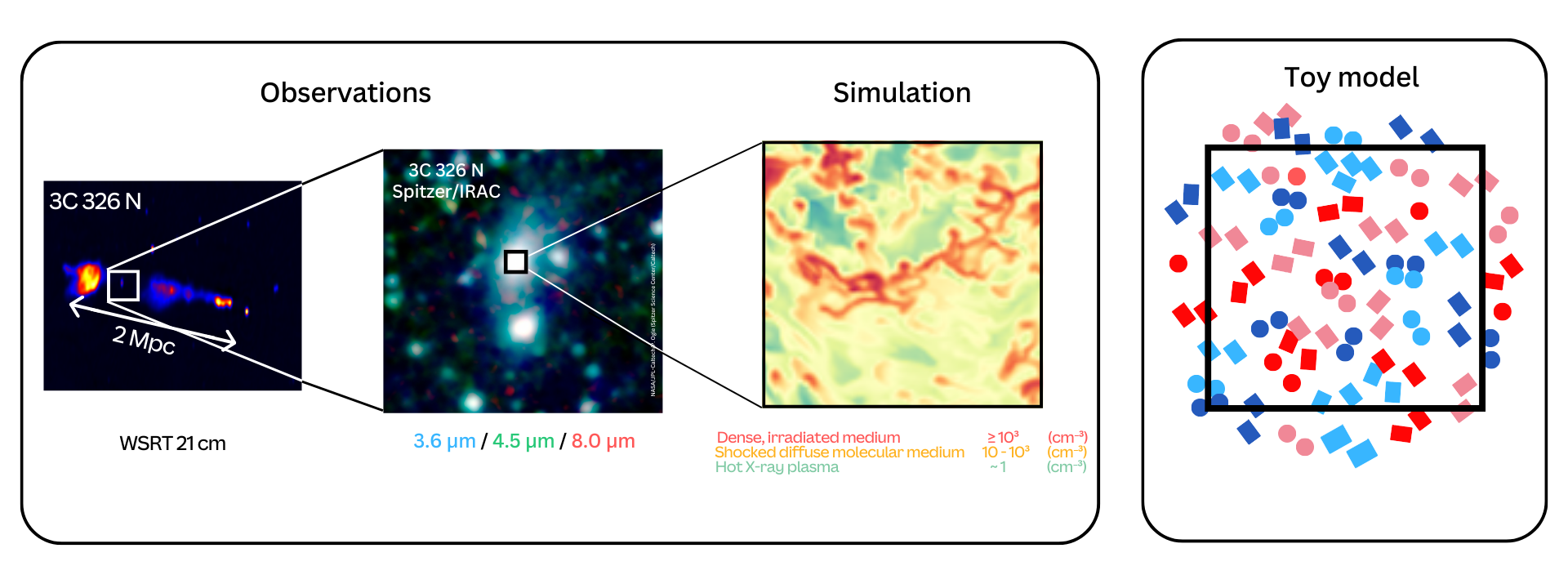}
        \caption{Schematic representation of the multiphase ISM in a galaxy. \textbf{Left:} radio emission observed in 3C~326~N with the Westerbork Synthesis Radio Telescope (WSRT) 21-cm \citep{Willis78} and \textit{Spitzer}/IRAC composite image \citep{Ogle07} at $3.6, 4.5,~{\rm and}~8.0~\mu{\rm m}$ showing the 3C~326~N galaxy (and its South companion). The $3.7''\times3.7''$ ($6.4\times6.4$~kpc) square (not-to-scale) represents the solid angle $\Omega_{\rm obs}$ over which line emissions are observed. \textbf{Center:}. Cutout showing the projected proton density from a numerical simulation of the interaction between a radio jet and the central part of a multiphase galactic disk \citep[from][]{Murthy22}. The different colors show the parsec-scale multiphase structure, with the dense irradiated medium in red ($\rm n_H \geq 10^3~cm^{-3}$), the shock diffuse gas in yellow (i.e., $\rm n_H \sim 10-10^3~cm^{-3}$), and the hot X-ray plasma in green ($\rm n_H \sim 1~cm^{-3}$). \textbf{Right:} simplified model used in this paper where the observational beam is supposed to encompass distributions of shock and PDR surfaces (see Fig.~\ref{fig1_2}).
        }
    \label{fig1_1}
\end{figure*}

The grid of models used in this paper is a subsample of the precomputed models obtained with the 1.5.4 version of the Meudon PDR code and publicly available on the InterStellar Medium DataBase service (ISMDB\footnote{Available at: \url{http://ism.obspm.fr/index.html}}). We consider PDR models at constant proton density only, with a left-side illumination scaling factor $\rm G_0 = 1$ or 10 and a right-side illumination scaling factor $\rm G_0 = 1$. The range of parameters explored here is shown in Table \ref{table:grid_params}.

\begin{table}
\caption{PDR and shock model parameters.}     
\label{table:grid_params}     
\centering 
\resizebox{\columnwidth}{!}{%
\begin{tabular}{lccc}
\hline\hline
 \textbf{PDR parameter} &  \textbf{Symbol} &  \textbf{Value} &  \textbf{Units}  \\
 \hline \\
 Proton density & $\rm n_H$ &  10,$10^2$,$10^3$,$10^4$ & $\rm cm^{-3}$ \\
 Radiation field & $\rm G_0$ & $1, 10$ & --- \\
 Total visual extinction & $\rm A{_{\rm v~max}}$ & $10$ & $\rm mag$ \\
 H$_2$ CR ionization rate & $\zeta_{\rm H_2}$ & $10^{-16}$ & $\rm s^{-1}$ \\
 \hline \\
 
\hline\hline
 \textbf{Shock parameter} &  \textbf{Symbol} &  \textbf{Value} &  \textbf{Units}  \\
 \hline \\
 Proton density & $\rm n_H$ &  10,$10^2$,$10^3$,$10^4$ & $\rm cm^{-3}$ \\
 Radiation field$^{\rm (a)}$ & $\rm G_0$ & $0, 1, 10$ & --- \\
 Initial visual extinction & $\rm A{_{\rm v}}$ & $1.0$ & $\rm mag$ \\
 Shock velocity & $\rm v_s$ & $1-40$~in steps of 1 & $\rm km~s^{-1}$ \\
 Magnetic parameter$^{\rm (b)}$ & $\rm b$ & $0.1,1.0$ & --- \\
 H$_2$ CR ionization rate & $\zeta_{\rm H_2}$ & $10^{-16}$ & $\rm s^{-1}$ \\
 \hline \\
\end{tabular}

} \\

{\small
{\raggedright \textbf{Notes.} \par}
{\raggedright $^{\rm (a)}$ Scaling factor of the standard ultraviolet radiation field of \cite{Mathis83}. \par}
{\raggedright $^{\rm (b)}$ Transverse magnetic field $\rm B_0 = b(n_H/cm^{-3})^{1/2}~\mu G$. \par}
}
\end{table}

\subsection{Grids of shock models}\label{sec:PDS_models}

The Paris-Durham shock code is a model designed to study the dynamical, thermal, and chemical evolution of molecular shocks propagating at a velocity $\rm v_s$ in a magnetized and irradiated environment with an initial proton density, $\rm n_H$, and an initial magnetic field strength set by a magnetic parameter $\rm b$. The geometry, the main assumptions, and the physical ingredients included in the code are described in \cite{Godard19}. In a nutshell, the model considers a 1D plane-parallel shock wave at steady-state traveling in a photo-dissociation region. The shock is assumed to be irradiated upstream by a semi-isotropic radiation field set to the standard interstellar radiation field (ISRF, \citealt{Mathis83}) and scaled with a parameter $\rm G_0$, and to be initially located at a position within the PDR set by a visual extinction $\rm A_v = 1.0$ \citep{Kristensen23}. Within this geometry, the model follows in a Lagrangian frame, the out-of-equilibrium evolution of a fluid particle during its trajectory from the pre-shock medium to the post-shock gas, starting from initial conditions at thermal and chemical equilibrium.

In addition to the chemical composition of the gas, the model solves the excitation of atomic and molecular species and the out-of-equilibrium evolutions of the first 150 ro-vibrational levels of H$_2$. In particular, the excitation of the fine structure levels of C$^+$ and O and of the ro-vibrational levels of H$_2$ are computed taking into account all the processes included in the Meudon PDR code and described in the previous section. The emerging line fluxes of all these species are finally calculated by integrating the emissivities over the entire shock up to the point where 99.9\% of the input mechanical energy is dissipated (see Eq. 1 of \citealt{Kristensen23}). This cutting point is chosen to prevent the inclusion of large amounts of cold post-shocked gas in the computation of line intensities and to ensure that emerging intensities mostly come from the reprocessing of mechanical energy input and not from other sources of energy (UV and IR fields, cosmic rays).

Although large grids of models obtained with the Paris-Durham shock code have been recently computed by \citet{Kristensen23} and are publicly available on the ISMDB service, we decide here to adopt a more refined set of shock velocities. The values of the parameters covered by the grid are shown in Table \ref{table:grid_params}. The shock velocity ranges between 1 and 40~km~s$^{-1}$ in steps of 1~km~s$^{-1}$.

\section{Ensembles of shocks and PDRs}\label{sec:distribution-shocks}

When an extragalactic object is observed, the instrumental beam captures the emission from all the phases of the ISM. As schematized in Fig.~\ref{fig1_1}, those include the hot and diffuse ionized gas, the diffuse and irradiated neutral media, and the dense molecular gas. In principle, all these phases contribute to the reprocessing of the radiative energy input originating from the AGN and the stars, and the mechanical energy input originating, for instance, from the interaction between the jet and the galactic disk. Since we focus on molecular lines and on typical tracers of the neutral interstellar medium, we neglect here the contribution of the hot ionized phase. Under this approximation, we develop a simplified model in which the reprocessing of the input UV radiative energy occurs in idealized 1D PDRs, and the reprocessing of the input mechanical energy occurs in idealized 1D shock structures.

\subsection{Toy model}\label{sec:toy_model}

As described in Appendix~\ref{sec:Lehman_toymodel}, we consider that an ensemble of shock and PDR surfaces, subtending total solid angles $\Omega_{\rm S}$ and $\Omega_{\rm P}$, are captured within an observational beam of solid angle $\Omega_{\rm obs}$ (see Fig.~\ref{fig1_1}). The cumulative fluxes originating from shock and PDR structures collectively contribute to the observed emission lines. For simplicity, the total solid angles $\Omega_{\rm S}$ and $\Omega_{\rm P}$ are assumed to follow 1D probability distribution functions denoted by $f_{\rm S}$ and $f_{\rm P}$. The shock distribution is supposed to depend on the shock velocity $\rm v_s$ and the PDR distribution is supposed to depend on the illumination factor $\rm G_0$, yielding the specific functional forms $f_{\rm S}({\rm v_s}) = d\Omega_{\rm S}/d{\rm v_s}$ and $f_{\rm P}({\rm G_0}) = d\Omega_{\rm P}/d{\rm G_0}$, respectively.

Considering that the continuum absorption due to the shock and PDR surfaces is negligible and that no cross-absorption occurs between the surfaces\footnote{Regardless of this assumption, the radiative transfer inside each shock and PDR structure is computed taking into account absorption processes.} (see Eqs. \ref{Eq-cond1} and \ref{Eq-cond2}), we derive a simple expression of the continuum subtracted line flux (see Eq. \ref{Eq-final}). It follows that the integrated intensity of each line (in $\rm erg~s^{-1}~cm^{-2}~sr^{-1}$) is given by
\begin{equation}
    I^{\rm m} = \frac{1}{\Omega_{\rm obs}} \left(\int_{\rm v_s} f_{\rm S}({\rm v_s}) I_{\rm s}({\rm v_s})d{\rm v_s} + \int_{\rm G_0} f_{\rm P}({\rm G_0}) I_{\rm p}({\rm G_0})d{\rm G_0} \right),
    \label{Eq-Imod}
\end{equation}
where $I_{\rm S}(\rm v_s)$ and $I_{\rm P}(\rm G_0)$ are the line integrated intensities emitted by one shock of velocity $\rm v_s$ and one PDR of illumination factor $\rm G_0$, respectively. The superscript `m' stands for model.

Note that the framework can be easily extended to other types of contributions to the line emission, such as XDRs and HII-region for instance. In such cases, the right-hand side of Eq.~\ref{Eq-Imod} can be partitioned into multiple components.
\subsection{Comparison between model and observations: distance minimization}\label{sec:fittingmethod}

This toy model can be used to interpret an ensemble of integrated intensities of unresolved emission lines. The confrontation of the model to the observations only requires defining a distance between the modeled and the observed intensities. Minimizing this distance leads to the characterization of the distributions of shocks and PDRs within the beam, in particular the number of shocks and PDRs and the distribution of shock velocities required to reproduce the observations.

Let's consider an ensemble of N emission lines. According to Eq.~\ref{Eq-Imod}, the modeled intensity of each line $j$ writes
\begin{equation}
    I_j^{\rm m} = \frac{1}{\Omega_{\rm obs}} \left( \int_{\rm v_s} f_{\rm S}({\rm v_s}) I_{j, \rm s}({\rm v_s})d{\rm v_s} + \int_{\rm G_0} f_{\rm P}({\rm G_0}) I_{j, \rm p}({\rm G_0})d{\rm G_0} \right).
    \label{eq:3}
\end{equation}
We define the distance between the modeled and the observed intensities of each line $j$ as the difference between the model prediction and the observation in log space
\begin{equation}
  {\rm d}_j = \left\{
     \begin{array}{@{}l@{\thinspace}l}
       0 & ~~ \text{if} ~~  I^{\rm o}_j - \frac{W_j}{2} \le I^{\rm m}_j \le I^{\rm o}_j + \frac{W_j}{2} \\
       \log_{10}(I^{\rm m}_j) - \log_{10}\left(I^{\rm o}_j - \frac{W_j}{2}\right)   & ~~ \text{if} ~~  I^{\rm m}_j < I^{\rm o}_j - \frac{W_j}{2} \\
       \\log_{10}(I^{\rm m}_j) - \log_{10}\left(I^{\rm o}_j + \frac{W_j}{2}\right)   & ~~ \text{if} ~~  I^{\rm m}_j > I^{\rm o}_j + \frac{W_j}{2}, \\

     \end{array}
   \right.
    \label{eq:4}
\end{equation}
where $I^{\rm o}_j$ is the observed intensity of line $j$ and $W_j$ is the associated one $\sigma$ uncertainty. The total quadratic distance between the model and the ensemble of observations is calculated as the sum of the quadratic distances of all emission lines
\begin{equation}
    {\rm d} = \sqrt{\sum_{j=1}^{N} {\rm d}^2_j}.
    \label{eq:4_distance}
\end{equation}
Eq.~\ref{eq:4_distance} is nothing more than the distance, in log scale, between the model and a hypercube in N dimensions defined by the observed intensities and their associated uncertainties. Such distance also bears a natural physical meaning. For instance, a distance of 1~dex implies that at most one observation is underestimated or overestimated by a factor of 10.

With this definition of the distance, the interpretation of an ensemble of atomic and molecular lines is performed using the following procedure:
\begin{enumerate}
  \item The 1D distribution functions $f_{\rm S}({\rm v_s})$ and $f_{\rm P}({\rm G_0})$ are assumed to have known functional forms. Based on the results of numerical studies (e.g., \citealt{Lehamnn16}), we adopt by default here an exponential distribution of shock velocities: $f_{\rm S}({\rm v_s}) = (\Omega_{\rm S}/\sigma_{\rm v_s})\exp(-{\rm v_s}/\sigma_{\rm v_s})$, where $\sigma_{\rm v_s}$ is the dispersion of shock velocities. For simplicity, we consider a Dirac delta distribution of PDRs illumination factor: $f_{\rm P}({\rm G_0}) = \Omega_{\rm P}\delta(\rm G_0)$.
  \item The modeled intensities are computed using Eq.~\ref{eq:3} over a wide range of the parameters $\Omega_{\rm S}$, $\sigma_{\rm v_s}$, and $\Omega_{\rm P}$ describing the distributions $f_{\rm S}$ and $f_{\rm P}$, assuming that all the other parameters (e.g., $\rm n_H$, b, see Table~\ref{table:grid_params}) are constants. The associated distances $\rm d(\Omega_{\rm S},\sigma_{\rm v_s},\Omega_{\rm P})$ between the model and the set of observations are computed with Eq.~\ref{eq:4_distance}.
  \item Global and local minima of the distance are finally searched for to identify the distributions $f_{\rm S}$ and $f_{\rm P}$ which fall the closest to the observational dataset.
\end{enumerate}
\subsection{Mechanical an UV-reprocessed luminosities}\label{sec:fittingmethod_luminosities}

The probability distribution functions $f_{\rm S}$ and $f_{\rm P}$ directly provide the amount of mechanical and radiative energies reprocessed by shocks and PDRs. The rate of dissipation of mechanical energy of a steady-state shock propagating at a velocity $\rm v_s$ in a medium with pre-shock mass density $\rho_0$ is 
\begin{equation}
\mathcal{L}_{\rm K}({\rm v_s}) = \frac{1}{2} \rho_0 {\rm v_s^3}\mathcal{A},
    \label{eq:5_0}
\end{equation} where $\mathcal{A}$ stands for the shock surface area (see Fig. \ref{fig1_2}). 
The total mechanical luminosity dissipated (in $\rm erg~s^{-1}~sr$) by a distribution of shocks at different velocities therefore writes
\begin{equation}
\mathcal{L}_{\rm K} = \frac{1}{2} \rho_0 \left(\int_{\rm v_s} f_{\rm S}({\rm v_s}) {\rm v^3_s} d{\rm v_s} \right){\rm D^2_A},
    \label{eq:5}
\end{equation} where ${\rm D_A}$ is the angular diameter distance.

As explained in Sect. \ref{sec:PDR_models}, a PDR is modeled as a plane-parallel cloud illuminated by a semi-isotropic ISRF with a fixed scaling factor on the backside and a varying radiation factor $\rm G_0$ on the front side. The integrated flux of the \cite{Mathis83} ISRF (in $\rm erg~cm^{-2}~s^{-1}$) from 912 to 2400 \AA~\citep{LePetit2006} implies that the UV-reprocessed power by one PDR is
\begin{equation}
\mathcal{L}_{\rm UV}({\rm G_0}) = 1.92\times10^{-3}~({\rm G_0 + 1.0})\mathcal{A}.
    \label{eq:6_0}
\end{equation} The total reprocessed UV-luminosity for an ensemble of PDRs that follow a 1D probability distribution $f_{\rm P}({\rm G_0})$, therefore writes
\begin{equation}
    \mathcal{L}_{\rm UV} = 1.92\times10^{-3}~({\rm G_0 + 1.0}) \left(\int_{\rm G_0} f_{\rm P}({\rm G_0}) d{\rm G_0}\right){\rm D^2_A}.
    \label{eq:6}
\end{equation}
\subsection{Shock and PDR masses}\label{sec:fittingmethod_mass}

The probability distribution functions $f_{\rm S}$ and $f_{\rm P}$ also directly provide the total masses of shocks and PDRs within the observational beam. The total mass of shocked material writes
\begin{equation}
    {\rm M}_{\rm S} = 1.4 {\rm m_H} \left(\int_{\rm v_s} f_{\rm S}({\rm v_s}) {\rm N_H(v_s)} d{\rm v_s} \right) {\rm D^2_A},
    \label{eq:shock_mass}
\end{equation} where $\rm m_H$ is the hydrogen mass and $\rm N_H(v_s)$ is the proton column density across a shock of velocity $\rm v_s$. The total mass of PDR material depends on the typical PDR size or equivalently their typical visual extinction $\rm A_v$. This mass writes
\begin{equation}
    {\rm M}_{\rm P} ({\rm A_v}) = 1.4 {\rm m_H} \left(\int_{\rm G_0} f_{\rm P}({\rm G_0}) d{\rm G_0} \right) {\rm N_H(A_v)} {\rm D^2_A},
    \label{eq:PDR_mass}
\end{equation} where $\rm N_{H}(A_v)$ is the proton column density evaluated at a given visual extinction $\rm A_v$.
\section{Application to the radio galaxy 3C~326~N}\label{sec:results}

Active galaxies are natural sources to investigate the importance of mechanical heating \citep{Begelman05}. As a test-case scenario, we therefore apply the framework developed in Sect. \ref{sec:distribution-shocks} to the radio galaxy 3C~326~N. This particular choice is motivated by the unique features of 3C~326~N, which include a notably low star formation rate \citep[$\sim 0.07~{\rm M_\odot~yr^{-1}}$]{Ogle10}, a substantial reservoir of molecular content \citep[$\sim 2\times 10^9~\rm{M_\odot}$]{Nesvadba11}, and the conspicuous presence of H$_2$ emission lines (see Appendix \ref{sec:3C326source}). The low star formation rate in this galaxy implies that the gas is weakly irradiated by the stellar UV radiation field and therefore favors the contributions of shocks in line emission \citep{Guillard12, Guillard15}. As described in Appendix \ref{sec:3C326source}, we consider a dataset composed of 14 emission lines which include \Hmol\ pure-rotational and ro-vibrational transitions, alongside [CII]~$158~\mu$m and [OI]~$63~\mu$m emission lines. The observed fluxes and uncertainties used for the distance minimization (see Eq. 3) are listed in Table \ref{table:1}.

The grids of PDR and shock models are described in Table \ref{table:grid_params}. Any observed emission line is assumed to result from the combination of PDRs and shocks within the observational beam\footnote{We do not consider the heating of the molecular gas by the X-ray radiation from the AGN. This is not a strong limitation, as \cite{Ogle10} showed that the H$_2$-to X-ray luminosity ratio ($> 0.6$) is so high that X-ray heating is not a primary driver of H$_2$ line emission.}. For simplicity, we consider that all PDRs and shocks have a unique and common density and pre-shock density, $\rm n_H$, and are illuminated by a unique and common UV radiation field characterized by a scaling factor $\rm G_0$. As explained in Sect. \ref{sec:modelling}, we also consider that shocks propagate in PDRs at a typical visual extinction\footnote{Note that this choice of typical visual extinction has little impact on the results presented here because the predictions of the shock models regarding H$_2$ emission are insensitive to $\rm A_v$ as long as the shocks run in a medium mostly molecular \citep{Kristensen23}.} $\rm A_v = 1.0$. Finally, we assume that all shocks propagate in a medium with a constant transverse magnetic field strength $\rm B_0$, scaled by the magnetization parameter b that follows the relation $\rm B_0 = b(n_H/cm^{-3})^{1/2}~\mu G$.

As a fiducial model, we adopt $\rm n_H = 10~cm^{-3}$, $\rm b = 0.1$, and $\rm G_0 = 10$. The choice of density comes from the exploration of the parameters and the comparison with the observations shown in the following sections. The choice of the illumination factor comes from the measurements of the UV emission \citep{Ogle10} and of the spectral SED of dust in the IR \citep{Guillard15} which suggest $\rm G_0 = 6^{+4}_{-2}$ and $\rm G_0 = 9 \pm 1$, respectively.

In this section, we first present the minimization process and the optimal shock and PDR distributions obtained for the fiducial model, and then show how this optimal distribution depends on the fixed parameters. In principle, the effects of all fixed parameters should be discussed. However, we find that both $\rm G_0$ and $\rm b$ have a minor impact on the distribution of shocks and PDRs required to explain the observations as long as $1 \leq \rm G_0 \leq 10$ and $\rm b \leq 1$. We will therefore limit our discussion to the effect of the gas density only.

\subsection{Likelihood analysis of the fiducial model}\label{ssec:param_likelihood}

\begin{figure*}
    \centering
     \includegraphics[width=0.95\textwidth,clip]
     {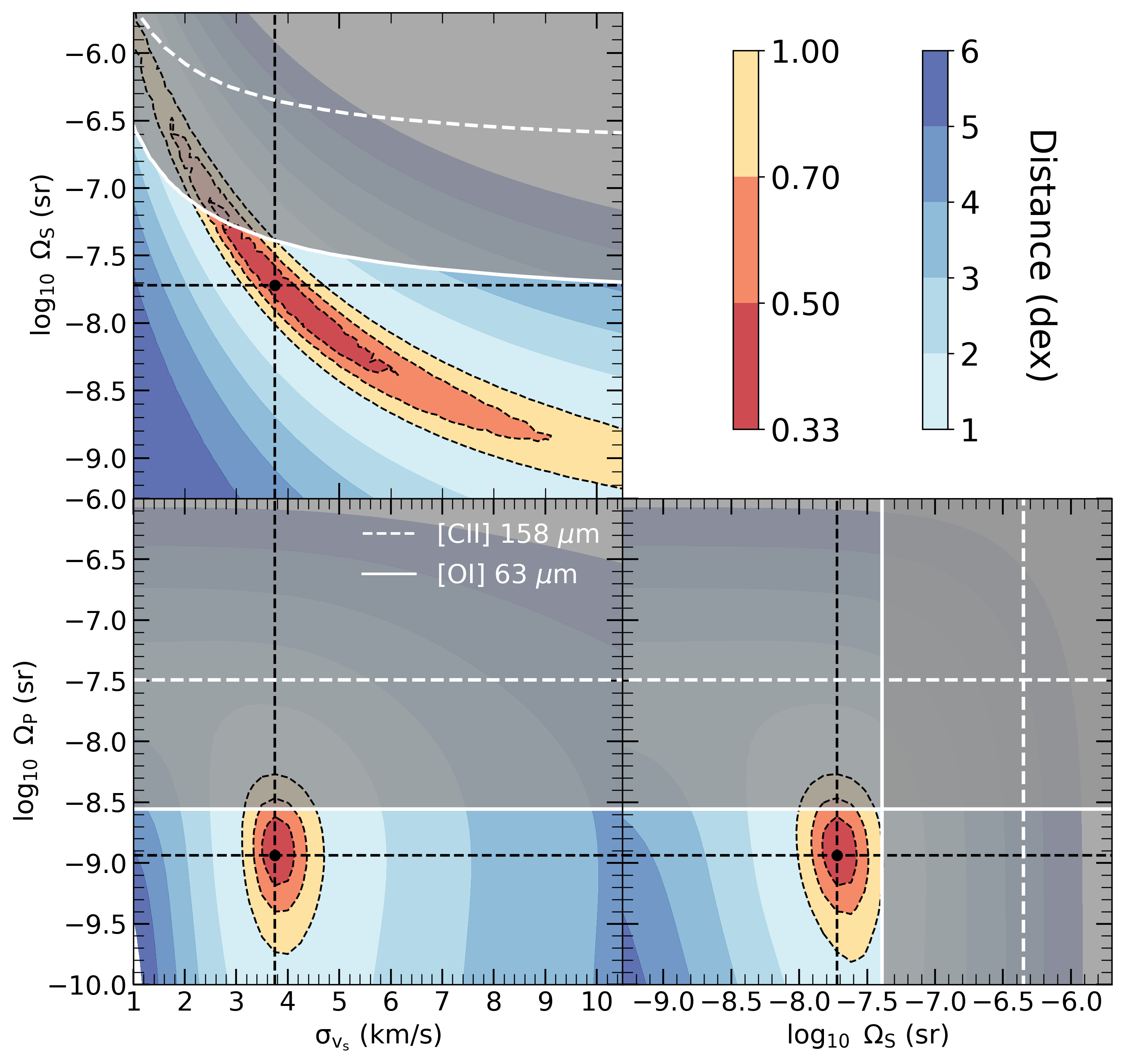}
    \caption{2D cutoff of the likelihood distributions of the main parameters describing the shock and the PDR distributions obtained for the fiducial model. The parameters are the total shock solid angle $\Omega_{\rm S}$, the dispersion of the shock velocities $\sigma_{\rm v_s}$, and the PDR total solid angle $\Omega_{\rm P}$. The color code represents the distance in dex units. The $0.5$, $0.7$, and $1.0$ dex limits are highlighted with dashed black lines. The straight black dashed lines and the black points indicate the position of the global minimum. Gray areas represent the regions where the assumption on the radiative transfer regarding cross absorption between surfaces breaks down. Note that the criterion is more stringent for the [OI]~$63~\mu$m line (gray area above the white solid line) than for the [CII]~$158~\mu$m line (gray area above the white dashed line) and is always fulfilled for H$_2$ lines (see Appendix~\ref{sec:Lehman_toymodel}).} 
    \label{fig0_2}
\end{figure*}

The results of the minimization procedure applied to the fiducial model are displayed in Fig. \ref{fig0_2} which shows the distance computed with Eqs. \ref{eq:4} and \ref{eq:4_distance} as a function of the parameters of the shock and PDR distributions, $\Omega_{\rm S}$, $\sigma_{\rm v_s}$, and $\Omega_{\rm P}$. This analysis highlights several key features.

First and foremost, we find that the fiducial model leads to distances as small as $0.33$~dex. As we show in Fig. \ref{figshockandPDR}, the value of the global minimum implies that the $14$ observational constraints, which include atomic and molecular emissions, are all reproduced within a factor smaller than $2$, a surprising result given the simplicity of the adopted distributions and 1D, plane-parallel geometry of the models. The global minimum is obtained for distributions of shocks and PDRs with total solid angles $\Omega_{\rm S} = 1.9\times10^{-8}~{\rm sr}$ and $\Omega_{\rm P} = 1.2\times10^{-9}~{\rm sr}$, and a shock velocity distribution characterized by $\sigma_{\rm v_s} \approx 4~{\rm km/s}$. This result implies that there are about $10$ shock surfaces per PDR surface in the observational beam. It also indicates that the number of shocks rapidly decreases with the shock velocity in agreement with the distribution of shock velocities derived in magnetohydrodynamic (MHD) simulations of interstellar turbulence \citep[see Fig. 9 of][]{Lehamnn16}. Interestingly, the optimal value of $\sigma_{\rm v_s} \approx 4~{\rm km/s}$ implies that the peak of dissipation of mechanical energy occurs for $\rm v_s = 3 \sigma_{\rm v_s} \sim 12~{\rm km/s}$ and that most of the dissipation is mediated by shocks with velocities between $\sim 5$ and $\sim 20~{\rm km/s}$.

Another main feature of this analysis is the uniqueness of the optimal solution. As suggested by Fig. \ref{fig0_2}, we find that the entire parameter space is characterized by only one global minimum, which occupies a finite volume in the parameter space, and no local minima. This absence of local minima indicates that there are no complex degeneracies between the parameters of the shock and PDR distributions. In particular, the large distances obtained for small values of $\Omega_{\rm S}$ or $\Omega_{\rm P}$ imply that both shocks and PDRs are required to explain the observational dataset and that no acceptable solution (with a distance smaller than one dex) can be obtained with a sole distribution of shocks or PDRs. Nevertheless, Fig. \ref{fig0_2} also reveals some degeneracy between $\Omega_{\rm S}$ and $\sigma_{\rm v_s}$: solutions with distances smaller than $0.5$~dex can be obtained for $\Omega_{\rm S}$ varying over one order of magnitude and $\sigma_{\rm v_s}$ over a factor of $2$. This degeneracy comes from the fact that the emission lines of H$_2$ mostly originate from shocks with a specific range of velocities (see Sect. \ref{sec:discussion}) whose numbers are set by both $\Omega_{\rm S}$ and $\sigma_{\rm v_s}$.

Fig. \ref{fig0_2} finally reveals the domain of validity of the radiative transfer algorithm. As detailed in Appendix \ref{sec:Lehman_toymodel}, the radiative transfer used in this work is valid only if Eqs. \ref{Eq-cond1} and \ref{Eq-cond2} are satisfied. The grey area displayed in Fig. \ref{fig0_2} shows the regions where these conditions break down for different emission lines. Remarkably, we find that the global minimum obtained for the fiducial model falls within a valid zone. This result therefore justifies, a posteriori, the fact of using a simplified radiative transfer to study the emission of ensembles of shocks and PDRs in galaxies. The simplified radiative transfer is, however, not valid for all values of $\Omega_{\rm S}$ and $\Omega_{\rm P}$. The most stringent constraints on the validity of the radiative transfer are due to the $63~\mu$m line of O and, to a lesser extent, to the $158~\mu$m line of C$^+$. This feature simply comes from the high opacities of these two lines across shock and PDR surfaces. In practice, the grey areas should be analyzed with a more sophisticated radiative transfer algorithm for [OI] emission (and if necessary [CII]) taking into account saturation effects induced by the cross-absorption between all the PDR and shock surfaces. However, such treatment would be superfluous. Indeed, as implied by the results shown in Sect.~\ref{ssec:result_energy}, these ranges of parameters lead to solutions in contradiction with the radiative and mechanical energy budgets of the galaxy.
\subsection{Impact of the medium's density}\label{ssec:result_density}

\begin{figure}
    \centering
     \includegraphics[width=0.46\textwidth,clip]{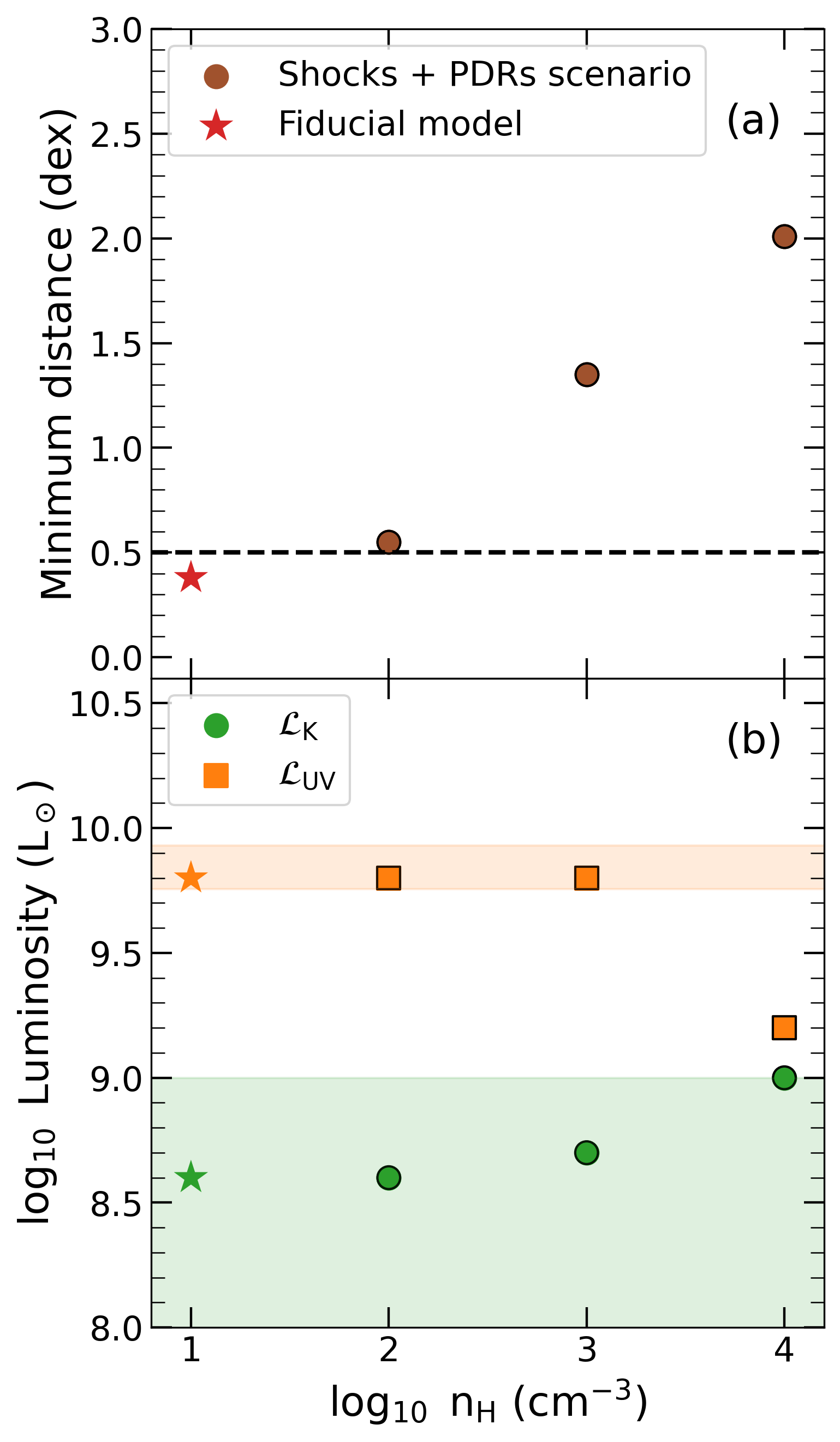}
    \caption{Solution for a combination of shocks and PDRs, assuming an exponential PDF. \textbf{Panel (a)}: distance versus the medium's density. The dash-line represents a 0.5 dex limit (i.e., a factor of three difference between observed and modeled line intensities) to guide the eye. \textbf{Panel (b)}: derived mechanical-reprocessed luminosity from Eq. \ref{eq:5} as well as the UV-reprocessed luminosity from Eq. \ref{eq:6} as a function of density are shown in green and orange colors, respectively. The green area represents $<1$\% of the estimated jet kinetic luminosity of 3C~326~N while the orange area represents the SED estimation of the reprocessed UV-luminosity (see Appendix \ref{ssec:SED}).}
    \label{figshockandPDR}
\end{figure}

\begin{figure*}[ht!]
    \centering
     \includegraphics[width=0.95\textwidth,clip]{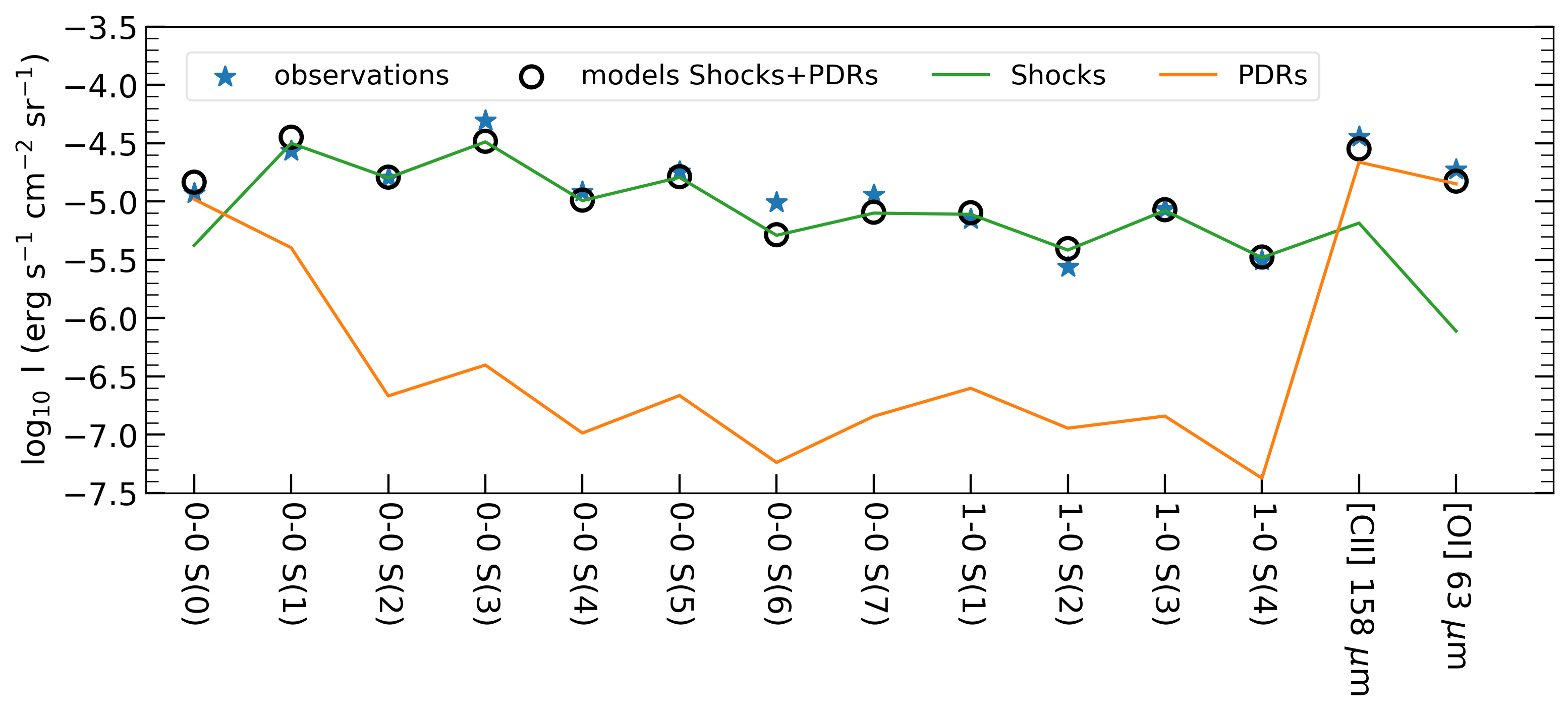}
    \caption{Solution for a combination of shocks and PDRs, assuming an exponential PDF. \textbf{Panel (c)}: solution for the fiducial model $\rm n_H = 10~cm^{-3}$, $\rm b = 0.1$, $\rm G_0 = 10$ shown in Fig. \ref{figshockandPDR} (a). Observations are presented as blue stars (see Table \ref{table:1}) and modeled intensities are shown as open black circles. Individual shock and PDR contributions are shown as green and orange solid lines, respectively. In the lower panel, the residuals are shown as black squares. The dash and solid lines give the $\rm \pm 0.5, 1~dex$ limits, respectively.}
    \label{figshockandPDR_1}
\end{figure*}

The impact of the density of the pre-shock gas and of the PDRs on the interpretation of observations is shown in Fig. \ref{figshockandPDR} (a) which displays the minimum distance between the observations and the distribution of shocks and PDRs models for a density ranging from $10$ to $\rm 10^4~cm^{-3}$. The smallest value of the minimum distance is obtained for $\rm n_H = 10~cm^{-3}$ which corresponds to the fiducial model. The corresponding intensities of the 14 emission lines used as constraints predicted by this fiducial model are shown Fig. \ref{figshockandPDR_1}.

As shown in Fig.~\ref{figshockandPDR_1}, the combination of shocks and PDRs reproduces all the observational lines by a factor smaller than 2. The H$_2$ lines from 0-0 S(1) to 1-0 S(4) are mostly emitted by shocks. Conversely, the 0-0 S(1) line of H$_2$, the $158~\mu$m line of C$^+$, and the $63~\mu$m line of O are mostly emitted by PDRs.

Fig. \ref{figshockandPDR} (a) shows that the minimum distance rapidly increases with the gas density. For instance, a minimum distance larger than $\sim 1.4$~dex is obtained for $\rm n_H = 10^3~cm^{-3}$, meaning that at most one of the observational lines is underestimated or overestimated by a factor $\sim 25$. This result is in contradiction with those currently presented in the literature. Fits of the excitation diagram of the H$_2$ pure-rotational transitions with shock models performed by \cite{Nesvadba10} for 3C~326~N lead to solutions with a density ranging between $10^3$ and $\rm 10^4~cm^{-3}$. These discrepancies come from the fact that our observational dataset contains additional constraints, in particular the [CII]~$158~\mu$m, and [OI]~$63~\mu$m lines. As shown in Fig.~\ref{fig_shockonly}, pure shock models cannot reproduce the [OI] line emission, even at low density.  Moreover, in  \cite{Nesvadba10}, a density of $\rm 10^4~cm^{-3}$ was needed to reproduce high-excitation lines, e.g. 0-0~S(7), but consequently, the 0-0~S(0) line was over-predicted by a factor of $\approx 2.5$. The fact that the [CII]~$158~\mu$m line originates from the diffuse gas is in agreement with \citet{Guillard15} who found similar results based on the observation of the [CII]-to-FIR ratio in 3C~326~N. It is also in line with \citealt{Pathak2024} who recently showed that the diffuse medium has a substantial contribution to the overall emission of the dust in main sequence galaxies.
\subsection{Reprocessed radiative and mechanical energies}\label{ssec:result_energy}

This new interpretation is valid only if the structures required to explain the observed intensities are in agreement with the energy budget of the source. This sanity check, which is often neglected, is paramount. The method described in Sect. \ref{sec:fittingmethod_luminosities} allows to compute the mechanical energy dissipated by shocks, $\mathcal{L}_{\rm K}$, as well as the UV-reprocessed luminosity in PDRs, $\mathcal{L}_{\rm UV}$. The resulting luminosities are shown in Fig. \ref{figshockandPDR} (b) as functions of the pre-shock and PDR densities and compared with the values estimated for 3C~326~N. For the fiducial model ($\rm n_H = 10~{\rm cm}^{-3}$, $\rm b = 0.1$, and $\rm G_0 = 10$), we derive a dissipated mechanical power of $\mathcal{L}_{\rm K} = 3.9\times10^8~{\rm L_\odot}$ and a UV-reprocessed luminosity of $\mathcal{L}_{\rm UV} = 6.3\times10^9~{\rm L_\odot}$. The mechanical energy dissipated by the shocks is therefore as high as $\sim 6$\% of the UV-reprocessed luminosity.

The UV luminosity predicted by the model corresponds to the radiation emitted by OB stars which is reprocessed by interstellar matter. This radiation is expected to be mostly absorbed by dust and re-emitted at infrared wavelengths. To obtain an independent estimate of the amount of UV radiation reprocessed, we perform a fit of the total SED of 3C~326~N, including 14 bands covering the UV-to-FIR range (see Appendix \ref{ssec:SED}), using the Bayesian code CIGALE\footnote{Available at: \url{https://cigale.lam.fr/}} \citep{Boquien19, Yang20, Yang22}. This treatment leads to a reprocessed UV luminosity $\rm L_{UV} = (7.1\pm1.4)\times10^9~L_\odot$. This estimate is in agreement with \cite{Lanz2016}, and is shown as an orange shaded region in Fig.~\ref{figshockandPDR} (b). This figure displays a remarkable agreement between the UV luminosity derived from the SED and that required to explain the atomic and molecular lines for gas densities ranging between $10$ to $\rm 10^3~{cm}^{-3}$. Interestingly, at $\rm n_H \gtrsim 10^4~cm^{-3}$, the UV-reprocessed luminosity obtained with the distribution of shocks and PDRs drops. This is due to the fact that at high densities, the 0-0~S(0) line is mainly emitted by low-velocity shocks instead of PDRs.

Estimation of the available mechanical energy is less straightforward. In radio-galaxies such as 3C~326~N, the interplay between the jet and the ISM results in a transfer of energy and momentum from large scales (the kpc-sized cavities inflated by the AGN-driven outflow, \citealt{Hardcastle20}) to smaller scales through a turbulent cascade, leading to an isotropization of the energy deposited by the AGN feedback, with turbulent velocity dispersions of the order of several hundreds of $\rm km~s^{-1}$ in the diffuse ionized phase \citep{Wittor2020}. This cascade leads to the formation of low-velocity shocks which may account for the dissipation of a substantial fraction of the turbulent mechanical energy in the molecular gas (e.g., \citealt{Lehamnn16, Park2019, Richard22}). According to simulations, approximately 20-30\% of the kinetic energy of the jet is deposited into the ISM, depending on the AGN power (e.g., \citealt{Mukherjee16}). Most of this energy is used to expand the gas along the jet and to drive gas outflows on sub-kpc scales \citep{Mukherjee18, Morganti23, Krause23}. Recent observations obtained with the JWST suggest that only a minor fraction of this energy (less than 1\%) feeds interstellar turbulence \citep{Pereira-Santaella2022}. \cite{Nesvadba10} estimated a power of the jet of $\rm \sim 10^{11}~L_\odot$ from radio observations. The upper limit computed as 1\% of this luminosity is shown as a green shaded area in Fig.~\ref{figshockandPDR} (b). For the fiducial model, the kinetic energy reprocessed in low-velocity shocks is found to be in agreement with the small fraction of the mechanical energy of the jet deposited in the ISM in 3C~326~N. Our results confirm that the dissipation of only a small fraction of the kinetic energy of the jet can account for the line emission of molecular hydrogen.
\subsection{Total masses and volume of shocks and PDRs}\label{ssec:result_mass}

The masses of shocks and PDRs in the fiducial model can be estimated using the prescription described in Sect.~\ref{sec:fittingmethod_mass}. The typical depth of PDRs is not known but since the UV-impinging power needs to be entirely reprocessed in infrared radiation, the associated visual extinction must be larger than 1. We assume two values of the visual extinction of PDRs, $\rm A_v = 1$ and $\rm A_v = 10$. Using Eqs. \ref{eq:shock_mass} and \ref{eq:PDR_mass}, we derive a total mass of shocks, ${\rm M}_{\rm S} = 7.8\times10^7~{\rm M}_\odot$, and a total mass of PDR, ${\rm M}_{\rm P} = 1.9\times10^8~{\rm M}_\odot$ (for $\rm A_v = 1$) and ${\rm M}_{\rm P} = 1.9\times10^9~{\rm M}_\odot$ (for $\rm A_v = 10$). This result implies that between 4 and 40\% of the total mass carried by the PDRs is shocked.

This estimation is in agreement with previous estimations of the molecular content of 3C~326~N. \cite{Ogle07} and \cite{Nesvadba10} derive a total mass of H$_2$ of $\rm 1.1\times10^9~M_\odot$ and $\rm 1.3-2.7\times10^9~M_\odot$, respectively, from the excitation diagram of H$_2$ observed in 3C~326~N. Similarly, \cite{Guillard15} estimate a total molecular mass of $\sim \rm 2\times10^9~M_\odot$ from the modeling of the FIR emission of dust grains and the emission of the CO(1-0) line.

\begin{figure*}[ht]
    \centering
     \includegraphics[width=0.95\textwidth,clip]{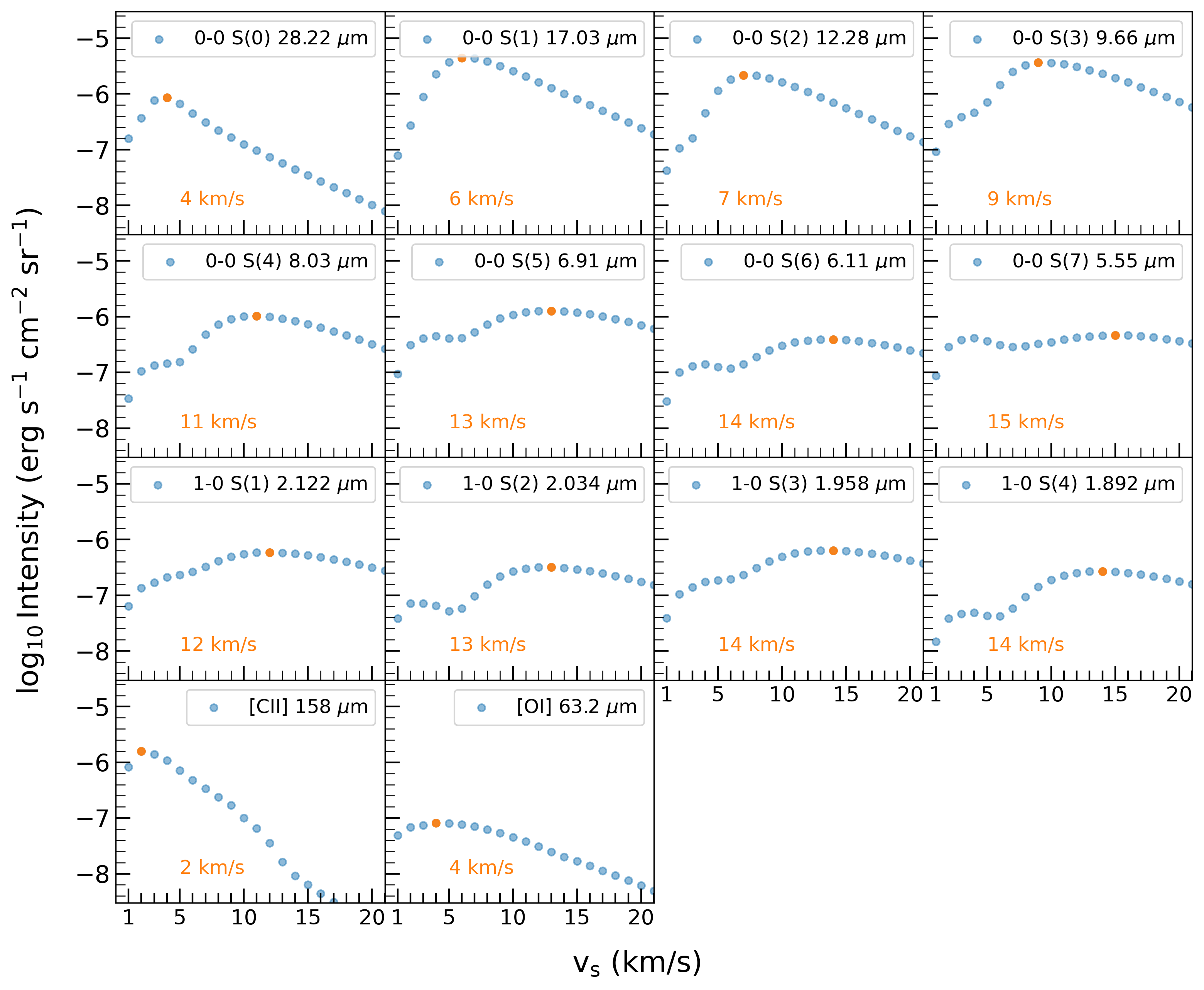}
    \caption{Contributions of the different velocities of the shock distribution to the total intensities emitted by shocks in the fiducial model. The dominant shock velocity, $\rm v_{s~dom}$, that contributes the most to powering each line is highlighted in orange. We display only a range between 1-20~km/s for readability, even though the models include velocities up to 40~km/s.} 
    \label{fig4_1}
\end{figure*}

The volume occupied by the PDRs can be derived from their total surface area and compared with the volume of the galaxy. For the fiducial model ($\rm n_H = 10~cm^{-3}$), we find that the volume of PDRs is comprised between $\rm 0.55~kpc^3$ (for $\rm A_v = 1$) and $\rm 5.5~kpc^3$ (for $\rm A_v = 10$). The volume of the galaxy is uncertain. Near-infrared IFU spectroscopy with the VLT/SINFONI instrument show that the H$_2$ ro-vibrational emission arises from a molecular disk of about $2$~kpc in diameter (see \citealt{Nesvadba11} footnote 1 page 3). The IR continuum and optical images show dust emission over $4$~kpc (see Fig. 3 in \citealt{Ogle10}). In addition, we estimate that the scale height of the galactic disk is between about $0.2$~kpc (see \citealt{Guillard15}) and $0.5$~kpc (see Fig. 3 in \citealt{Ogle10}). This implies that the molecular gas is contained in a volume between $5$ and $\rm 50~kpc^3$. These estimations show that the models with densities $\rm n_H \geqslant 10~cm^{-3}$ are coherent because they lead to a volume filling factor smaller than $1$. Given the results shown in Fig. \ref{figshockandPDR} one could wonder why we do not explore gas at densities lower than $\rm 10~cm^{-3}$. However, this additional constraint shows that such models are ruled out because the molecular gas would occupy a volume larger than that of the galaxy.
\subsection{Prediction of the CO(1-0) emission}\label{ssec:fitting_CO}

The observed CO(1-0) emission line intensity given in Table \ref{table:0_1} is not included as a constraint in our model. Low-J CO emission are known to arise from the cold diffuse component of molecular gas \citep{Mingozzi18}. Given that the total mass of shocks is much smaller than that of PDRs, it is therefore expected that the low-J lines of CO are mostly emitted by PDRs.

The distribution of PDRs obtained in the fiducial model leads to a flux of the CO (1-0) line of $6.5 \times 10^{-18}$~erg~s$^{-1}$~cm$^{-2}$, a value in fair agreement, yet slightly larger than that observed in 3C~326~N ($3.5 \times 10^{-18}$~erg~s$^{-1}$~cm$^{-2}$). This result reinforces our interpretation of this galaxy in which the 0-0 S(0), [CII]~$158~\mu$m, and [OI]~$63~\mu$m are mostly emitted by PDRs, while the remaining H$_2$ lines are mostly emitted by shocks.

\section{Model predictions for  JWST observations}\label{sec:discussion}
\subsection{Impact of shock velocities to the \texorpdfstring{H$_2$}{lines} lines}\label{ssec:discussion_shocktype}

From the fitting of observed line emission in 3C~326~N  (Fig. \ref{figshockandPDR}), it is clear that \Hmol\ lines are mainly powered by shocks, except the lowest excitation 0-0~S(0) line where the PDRs contribution dominate. For the fiducial model ($\rm n_H = 10~{\rm cm}^{-3}$, $\rm b = 0.1$, and $\rm G_0 = 10$, see Sect.~\ref{sec:results}), we now investigate, amongst the distribution of shocks, which shock velocities are the dominant contributors to the emission of \Hmol\ lines.

\begin{figure}
    \centering
     \includegraphics[width=0.47\textwidth,clip]{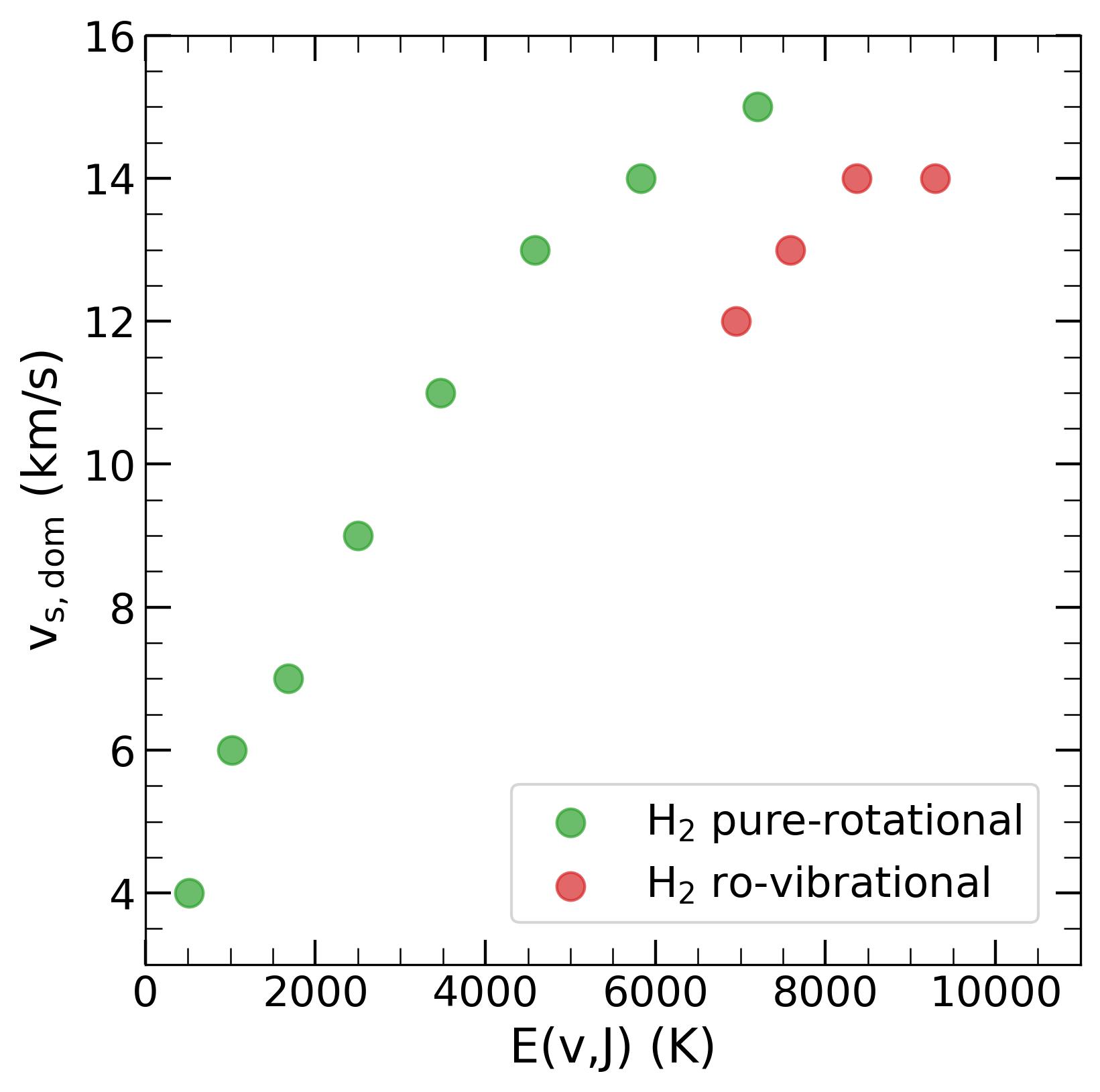}
    \caption{Shock velocity that contributes the most to the H$_2$ line emissions as a function of the upper-level energy (orange points in Fig.~\ref{fig4_1}). H$_2$ pure-rotational (green) and ro-vibrational (red) transitions are presented. $\rm v_{s, dom}$ is derived for the fiducial model $\rm n_H = 10~cm^{-3}$, $\rm b = 0.1$, and $\rm G_0 = 10$.}
    \label{fig3_app}
\end{figure}

Fig.~\ref{fig4_1} displays the modeled shock integrated line intensities for the suite of 14 emission lines as a function of the shock velocity. A summary of the dominant velocity as a function of the upper-level energy of the H$_2$ lines is presented in Fig.~\ref{fig3_app}. For the lines accessible with JWST observations, i.e.  0-0 S(1) to the 0-0 S(7), and the ro-vibrational 1-0~S(1) to 1-0~S(4), we show that the dominant shock velocities are remarkably low, in the range $\rm v_{s, dom} = 6~km/s$ to $\rm 15~km/s$. 
The typical length scales over which those shocks dissipate energy is of the order of 0.001--0.01~pc \citep{Kristensen23}. As previous studies pointed out, this suggests that a (small) fraction of the feedback kinetic energy cascades to very small scales in the molecular phase, powering the \Hmol\ line emission \citep{Nesvadba10}.

On a fairly more technical side, the shocks responsible for the \Hmol\ line emission are J-type (J for Jump) shocks \citep[see][for a discussion of the shock model grid, and in particular their Fig.~6]{Kristensen23}. Those shocks are strong emitters of ro-vibrational emission lines, contrary to C-type (C for Continuous) shocks, which were favored in \cite{Nesvadba10} with fewer observational constraints. Therefore, JWST/NIRSpec observations of 3C~326~N will be a stringent test of our model predictions.

\subsection{Impact of the distribution of shock velocities} \label{ssec:discussion_PDFS}

The impact of the functional form of the distribution of shock velocities is presented in Appendix~\ref{ssec:GaussianSol} where we show the comparison between the model and the observations assuming a Gaussian distribution for $f_{\rm S}$. Appendix~\ref{ssec:GaussianSol}, and in particular Figs.~\ref{fig0_2_app} and~\ref{shockpdr_app}, show that a Gaussian distribution leads to the same conclusions regarding the media responsible for the different line emissions and the energy budget required to reproduce the observations. In fact, Fig. ~\ref{fig0_2_app} reveals that assuming a more complex distribution of shock velocities only leads to additional degeneracies. For instance, we find that a narrow distribution of shock velocities centered at $\mu_{\rm v_s} = 12$~km~s$^{-1}$ leads to the same distance as a broad distribution centered at $\mu_{\rm v_s} = 3$~km~s$^{-1}$. This implies that the current ensemble of observations is insufficient to discriminate different distributions of shock velocities as long as the shocks that dominate the rovibrational emission of H$_2$ (between 6 and 15 km s$^{-1}$, see Fig.~\ref{fig4_1}) are included.

\subsection{\texorpdfstring{H$_2$}{line predictions for JWST} line predictions for JWST}\label{ssec:discussion_shockPDR}

\begin{figure*}
    \centering
     \includegraphics[width=0.99\textwidth,clip]{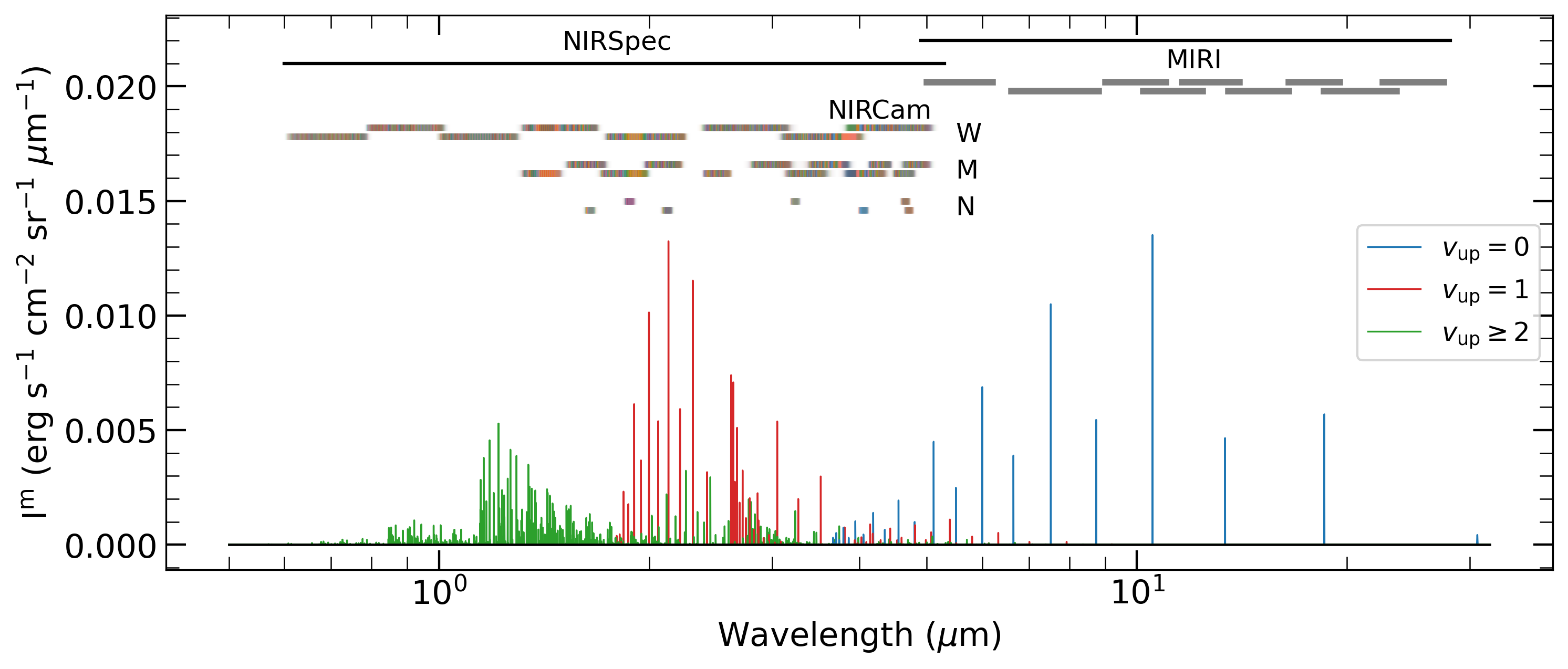}
    \caption{Predicted synthetic spectrum of H$_2$ in 3C~326~N obtained for the fiducial model ($\rm n_H = 10~{cm}^{-3}$, $\rm b = 0.1$, and $\rm G_0 = 10$). The specific intensities of the strongest $1000$ emission lines are shown as functions of the redshifted wavelength for $\rm z = 0.089$. We apply the relations described in Sect. \ref{ssec:discussion_shockPDR} for the resolving power of the NIRSpec and MIRI instruments. Outside the validity wavelength range, we fix the resolving power to a value of $\lambda/{\Delta\lambda} = 2500$. The NIRSpec and MIRI bandwidths correspond to the black solid lines. The wide-, medium-, and narrow-band filters for NIRCam and MIRI are shown as gray and color bars. Pure rotation lines are indicated in blue while ro-vibrational transitions are shown in red and green. 
    }
    \label{fig_spectrum}
\end{figure*}

\begin{figure*}[h!]
    \centering
     \includegraphics[width=0.99\textwidth,clip]{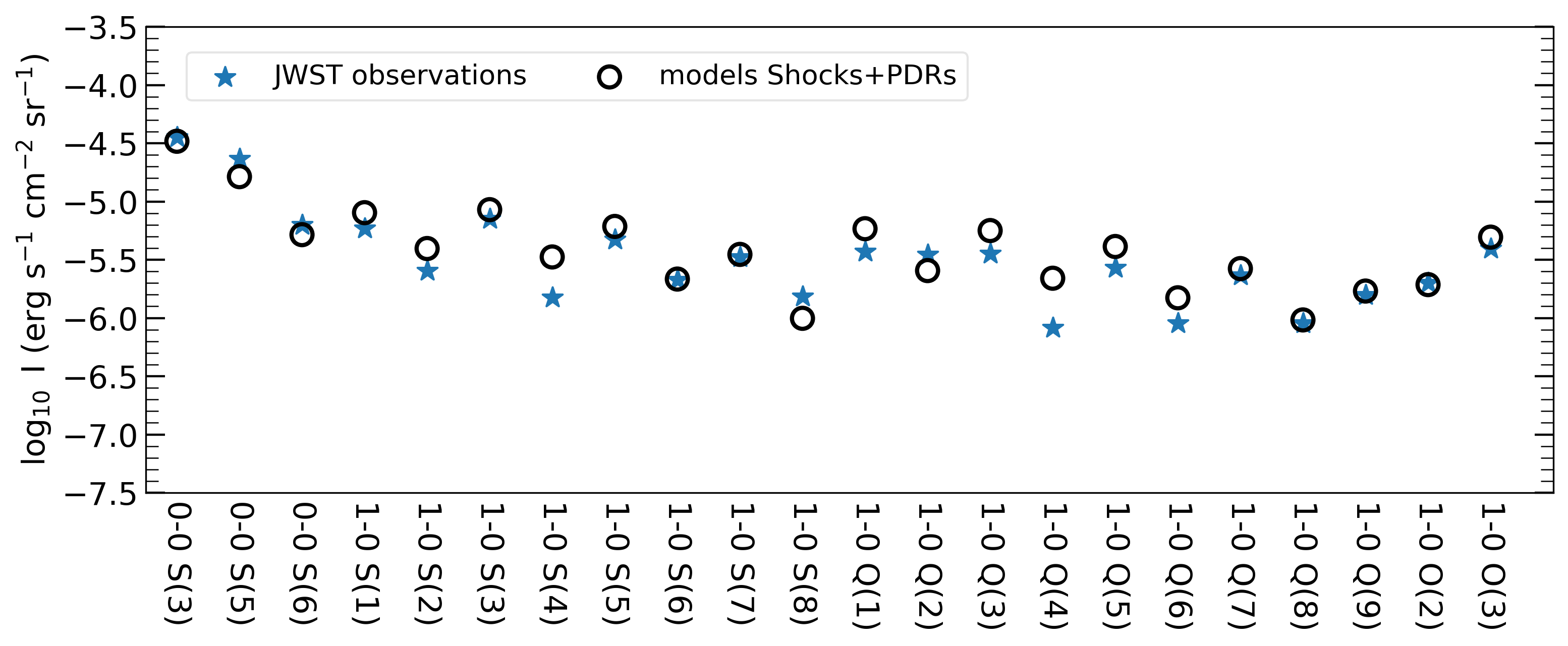}
    \caption{Comparison of the intensities of the H$_2$ lines predicted by the fiducial model with the new JWST data observed by \citealt{Leftley24}, which are corrected here by a factor $(1+z)^3$ (see Eq.~\ref{Eq-redshift}).  Observations are shown as blue stars and the model predictions as opened black circles. We stress that the model is obtained by fitting the \textit{Spitzer} and VLT/SINFONI data, and not the JWST data.}
    \label{fig_spectrum_JWST}
\end{figure*}

The Paris-Durham shock and the Meudon PDR codes solve for the populations of a large number of \Hmol\ levels (see Sect.~\ref{sec:modelling}) and therefore provide a predicted spectrum over the entire wavelength range observed by the JWST instruments. Fig.~\ref{fig_spectrum} displays the spectrum predicted by the fiducial model for the strongest 1000 lines of H$_2$, accounting for the different spectral resolutions of the NIRSpec and MRS spectrographs. We fit the resolving power of the G235H NIRSpec disperser\footnote{We use wavelength calibration data from \url{https://jwst-docs.stsci.edu/jwst-near-infrared-spectrograph/nirspec-instrumentation/nirspec-dispersers-and-filters}} over the wavelength range $1.7-3.1~\mu{\rm m}$, finding the resolving power relation $R(\lambda) = 1225\lambda - 185$. For the MIRI instrument, we use the relation $R(\lambda) = 4603 - 128\lambda$ reported in \cite{Argyriou2023} valid over the wavelength range $5-28~\mu{\rm m}$. Outside the validity range in wavelength of these relations, we use a uniform resolving power of $\lambda/\Delta\lambda = 2500$. Note that, although displayed on the synthetic spectrum, the ground-state 0-0~S(0) emission line is not observable with MIRI/MRS, because its sensitivity drops above 26.5~\mum.

We predict that the 3C~326~N spectrum should be dominated by ro-vibrational lines with $\upsilon_{\rm up} = 1$ around $2-3~\mu{\rm m}$, as well as the strong 0-0~S(3) and 0-0~S(5) mid-infrared lines. The forest of near-infrared $\upsilon_{\rm up} \geq 1$ ro-vibrational lines is thus predicted to be a major cooling channel for the dissipation of kinetic energy.

Luckily, it came to our attention, while this paper was under peer review, that new data obtained with the NIRSpec and MIRI instruments aboard the JWST had just been published \citep{Leftley24}. These data include 19 H$_2$ lines encompassing S, O, and Q series, as well as the pure-rotational lines 0-0~S(3), 0-0~S(5), and 0-0~S(6). It offered a unique and timely opportunity to test the predictive power of the model. The predictions of the fiducial model were taken as they were and directly compared to these new JWST data. Fig. \ref{fig_spectrum_JWST} displays the remarkable results that the fiducial model reproduces all the lines observed with the JWST within a factor of about 2. Also, as predicted by Fig. \ref{fig_spectrum}, the number of ro-vibrational lines present in the NIRSpec spectrum is abundant and is consistent with an emission powered by the reprocessing of mechanical energy from shocks.
\section{Summary and Conclusions}\label{sec:conclusions}

This paper presents a new framework to interpret atomic and molecular emission lines observed in extragalactic sources. Emission lines collected within an observational beam are assumed to arise from distributions of shocks and PDRs. The comparison of the observations with the predictions of the model leads to the physical properties (density, UV illumination, ...) of the emitting gas and provides the mechanical and radiative energy budget of the source. As a test case, we selected the radio galaxy 3C~326~N, a unique object with low SFR, large molecular content, and strong IR H$_2$ line emission. A total of 14 emission lines including eight H$_2$ rotational lines from 0-0 S(0) to 0-0 S(7) and four ro-vibrational lines from 1-0 S(1) to 1-0 S(4), as well as [CII]~$158~\mu$m and [OI]~$63~\mu$m emissions are used as observational constraints.

The framework developed in this paper is an oversimplification of the gas distribution in galaxies. Because the ISM is highly turbulent and contains a distribution of irradiation sources, both the density and $\rm G_0$ should spread over a wide range of values. Indeed, this diversity of ISM environments is obtained in numerical simulations (e.g., \citealt{Mukherjee18, Murthy22}) and derived from the analysis of continuum emission (e.g., \citealt{Pathak2024}). We designed the model and the underlying distributions to reduce the number of free parameters. This model is not a representation of the actual state of the gas but rather describes the mean physical conditions of the emitting medium. The remarkable result is that this oversimplified picture already leads to predictions in excellent agreement with the observations and fulfills the constraints on the mechanical and radiative energy budgets. A unique minimum is found with a small degree of degeneracies between the parameters (namely $\Omega_{\rm S}$ and $\sigma_{\rm v_s}$). This implies that adopting a distribution of density or $\rm G_0$, while more realistic, would only lead to additional degeneracies. The key results obtained within this framework are summarized as follows.

\begin{enumerate}
\item The analysis applied to 3C~326~N shows that the ensemble of observational constraints requires the combined contributions of both a distribution of shocks and a distribution of PDRs within the observational beam.
\item The predictions of the model are found to be highly dependent on the density of the gas and weakly dependent on the other parameters (the ambient radiation field and the magnetic field). Surprisingly, we find that only shocks and PDRs at low density ($\rm n_H = 10 - 100~cm^{-3}$) can reproduce all the observational constraints. This is probably the most salient result of this study. It implies that both the mechanical and UV energy inputs are preferentially reprocessed in the low-density medium of this galaxy.
\item The optimal solution is obtained for $\rm n_H = 10~cm^{-3}$. The likelihood, estimated from the ratios between the observed and the predicted intensities, is characterized by only one global minimum and no local minima. This global minimum provides constraints on the total solid angle of PDRs, the total solid angle of shocks, and the range of shock velocities required to explain the observations. However, the suite of spectral lines used in this work is found to be insufficient to set apart the predictions obtained for an exponential or a Gaussian probability distributions of shock velocities.
\item The H$_2$ 0-0~S(0)~$28~\mu$m, [CII]~$158~\mu$m, and [OI]~$63~\mu$m lines are mostly emitted by PDRs. The resulting UV-reprocessed luminosity, $\mathcal{L}_{\rm UV} = 6.3\times10^9~{\rm L_\odot}$, is found to be in remarkable agreement with the observed IR-luminosity.
\item Conversely, the rest of the H$_2$ lines from the 0-0~S(1) to the 1-0~S(4) are found to be emitted by low-velocity shocks ($\rm 5 < v_s < 20$~km~s$^{-1}$). The resulting mechanical luminosity dissipated by shocks, $\mathcal{L}_{\rm K} = 3.9\times10^8~{\rm L_\odot}$, is less than 1\% of the estimated jet kinetic power, in agreement with the predictions of MHD simulations or recent JWST observations.
\item The total mass of shocks is ${\rm M}_{\rm S} = 7.8\times10^7~{\rm M}_\odot$ and the total mass of PDRs is ${\rm M}_{\rm P} = 1.9\times10^9~{\rm M}_\odot$. This budget is in line with previous estimations of the molecular content in 3C~326~N and implies that about 4\% of the total mass carried by PDRs is shocked.
\item Finally, the predicted synthetic H$_2$ spectrum shows prominent ro-vibrational emission lines, originating from all the vibrational levels of H$_2$. This synthetic spectrum is found to be in excellent agreement with new JWST observations of 3C~326~N, underlining the ability of the model to make accurate predictions.
\end{enumerate}

The new interpretative framework is applied here to a specific source to test the validity and relevance of the underlying method. In a companion paper, we will extend this methodology to a larger sample of objects (see e.g. Fig.~7 in \citealt{Guillard12} and Fig.~12 in \citealt{Ogle23}), such as star-forming galaxies, AGN, radio, or elliptical galaxies and extract the radiative and mechanical energy budget of these sources from atomic and molecular line emissions.

The fact that most of the reprocessing of the input mechanical and radiative powers occurs in the diffuse gas in 3C~326~N is of particular importance. It remains to be seen whether this result holds for a larger sample of galaxies, which we will explore in a companion paper. It suggests that this methodology should be applied to ancillary interpretations of extragalactic sources. Previous studies \citep{Meijerink13, Mingozzi18, Esposito22} show that PDR and XDR models at high densities (e.g., $\rm n_H \sim 10^4~cm^{-3}$) could explain the CO, H$_2$, and [CII]~$158~\mu$m emission lines observed towards a few extragalactic sources. It would therefore be interesting to apply our methodology to those sources and estimate whether a combined distribution of shocks, PDRs, and XDRs at low densities could account for the observations with an acceptable energy budget. It is indeed timely because JWST is already observing  H$_2$ and fine-structure emission lines in extragalactic sources, in particular in AGN \citep{Pereira-Santaella2022, Buiten2023}.

\begin{acknowledgements}
We are grateful to the referee for their thorough reading and their insightful comments. The research leading to these results has received funding from the European Research Council, under the European Community’s Seventh framework Programme, through the Advanced Grant MIST (FP7/2017-2022, No 742719). The grid of simulations used in this work has been run on the computing cluster Totoro of the ERC MIST, administered by MesoPSL. We would also like to acknowledge the support from the Programme National “Physique et Chimie du Milieu Interstellaire” (PCMI) of CNRS/INSU with INC/INP co-funded by CEA and CNES. PG would like to thank the Institut Universitaire de France, the Centre National d'Etudes Spatiales (CNES), and the "Programme National de Cosmologie and Galaxies" (PNCG)  of CNRS/INSU, with INC/INP co-funded by CEA and CNES,  for their financial supports. 
\end{acknowledgements}

\bibliographystyle{aa}  
\bibliography{aa} 

\begin{thebibliography}{70}
\expandafter\ifx\csname natexlab\endcsname\relax\def\natexlab#1{#1}\fi

\bibitem[{{Allen} {et~al.}(2008){Allen}, {Groves}, {Dopita}, {Sutherland}, \&
  {Kewley}}]{Allen2008}
{Allen}, M.~G., {Groves}, B.~A., {Dopita}, M.~A., {Sutherland}, R.~S., \&
  {Kewley}, L.~J. 2008, \apjs, 178, 20

\bibitem[{{Almeida} {et~al.}(2023){Almeida}, {Anderson},
  {Argudo-Fern{\'a}ndez}, {Badenes}, {Barger}, {Barrera-Ballesteros}, {Bender},
  {Benitez}, {Besser}, {Bird}, {Bizyaev}, {Blanton}, {Bochanski}, {Bovy},
  {Brandt}, {Brownstein}, {Buchner}, {Bulbul}, {Burchett}, {Cano D{\'\i}az},
  {Carlberg}, {Casey}, {Chandra}, {Cherinka}, {Chiappini}, {Coker}, {Comparat},
  {Conroy}, {Contardo}, {Cortes}, {Covey}, {Crane}, {Cunha}, {Dabbieri},
  {Davidson}, {Davis}, {de Andrade Queiroz}, {De Lee}, {M{\'e}ndez Delgado},
  {Demasi}, {Di Mille}, {Donor}, {Dow}, {Dwelly}, {Eracleous}, {Eriksen},
  {Fan}, {Farr}, {Frederick}, {Fries}, {Frinchaboy}, {G{\"a}nsicke}, {Ge},
  {Gonz{\'a}lez {\'A}vila}, {Grabowski}, {Grier}, {Guiglion}, {Gupta}, {Hall},
  {Hawkins}, {Hayes}, {Hermes}, {Hern{\'a}ndez-Garc{\'\i}a}, {Hogg},
  {Holtzman}, {Ibarra-Medel}, {Ji}, {Jofre}, {Johnson}, {Jones}, {Kinemuchi},
  {Kluge}, {Koekemoer}, {Kollmeier}, {Kounkel}, {Krishnarao}, {Krumpe},
  {Lacerna}, {Lago}, {Laporte}, {Liu}, {Liu}, {Liu}, {Lopes}, {Macktoobian},
  {Majewski}, {Malanushenko}, {Maoz}, {Masseron}, {Masters}, {Matijevic},
  {McBride}, {Medan}, {Merloni}, {Morrison}, {Myers}, {M{\'e}sz{\'a}ros},
  {Negrete}, {Nidever}, {Nitschelm}, {Oravetz}, {Oravetz}, {Pan}, {Peng},
  {Pinsonneault}, {Pogge}, {Qiu}, {Ramirez}, {Rix}, {Fern{\'a}ndez Rosso},
  {Runnoe}, {Salvato}, {Sanchez}, {Santana}, {Saydjari}, {Sayres},
  {Schlaufman}, {Schneider}, {Schwope}, {Serna}, {Shen}, {Sobeck}, {Song},
  {Souto}, {Spoo}, {Stassun}, {Steinmetz}, {Straumit}, {Stringfellow},
  {S{\'a}nchez-Gallego}, {Taghizadeh-Popp}, {Tayar}, {Thakar}, {Tissera},
  {Tkachenko}, {Hernandez Toledo}, {Trakhtenbrot}, {Fern{\'a}ndez-Trincado},
  {Troup}, {Trump}, {Tuttle}, {Ulloa}, {Vazquez-Mata}, {Vera Alfaro},
  {Villanova}, {Wachter}, {Weijmans}, {Wheeler}, {Wilson}, {Wojno}, {Wolf},
  {Xue}, {Ybarra}, {Zari}, \& {Zasowski}}]{Almeida2023}
{Almeida}, A., {Anderson}, S.~F., {Argudo-Fern{\'a}ndez}, M., {et~al.} 2023,
  \apjs, 267, 44

\bibitem[{{Argyriou} {et~al.}(2023){Argyriou}, {Glasse}, {Law}, {Labiano},
  {{\'A}lvarez-M{\'a}rquez}, {Patapis}, {Kavanagh}, {Gasman}, {Mueller},
  {Larson}, {Vandenbussche}, {Glauser}, {Royer}, {Dicken}, {Harkett},
  {Sargent}, {Engesser}, {Jones}, {Kendrew}, {Noriega-Crespo}, {Brandl},
  {Rieke}, {Wright}, {Lee}, \& {Wells}}]{Argyriou2023}
{Argyriou}, I., {Glasse}, A., {Law}, D.~R., {et~al.} 2023, \aap, 675, A111

\bibitem[{{Begelman} \& {Ruszkowski}(2005)}]{Begelman05}
{Begelman}, M.~C. \& {Ruszkowski}, M. 2005, Philosophical Transactions of the
  Royal Society of London Series A, 363, 655

\bibitem[{{Benson}(2010)}]{Benson2010}
{Benson}, A.~J. 2010, \physrep, 495, 33

\bibitem[{{Boquien} {et~al.}(2019){Boquien}, {Burgarella}, {Roehlly}, {Buat},
  {Ciesla}, {Corre}, {Inoue}, \& {Salas}}]{Boquien19}
{Boquien}, M., {Burgarella}, D., {Roehlly}, Y., {et~al.} 2019, \aap, 622, A103

\bibitem[{{Bruzual} \& {Charlot}(2003)}]{BC03}
{Bruzual}, G. \& {Charlot}, S. 2003, \mnras, 344, 1000

\bibitem[{{Buiten} {et~al.}(2023){Buiten}, {van der Werf}, {Viti}, {Armus},
  {Barr}, {Barcos-Mu{\~n}oz}, {Evans}, {Inami}, {Linden}, {Privon}, {Song},
  {Rich}, {Aalto}, {Appleton}, {B{\"o}ker}, {Charmandaris}, {Diaz-Santos},
  {Hayward}, {Lai}, {Medling}, {Ricci}, \& {U}}]{Buiten2023}
{Buiten}, V.~A., {van der Werf}, P.~P., {Viti}, S., {et~al.} 2023, arXiv
  e-prints, arXiv:2312.01945

\bibitem[{{Carnall} {et~al.}(2018){Carnall}, {McLure}, {Dunlop}, \&
  {Dav{\'e}}}]{Carnall2018}
{Carnall}, A.~C., {McLure}, R.~J., {Dunlop}, J.~S., \& {Dav{\'e}}, R. 2018,
  \mnras, 480, 4379

\bibitem[{{Casey} {et~al.}(2014){Casey}, {Narayanan}, \& {Cooray}}]{Casey2014}
{Casey}, C.~M., {Narayanan}, D., \& {Cooray}, A. 2014, \physrep, 541, 45

\bibitem[{{Chabrier}(2003)}]{Chabrier2003}
{Chabrier}, G. 2003, \pasp, 115, 763

\bibitem[{{Charlot} \& {Fall}(2000)}]{CF00}
{Charlot}, S. \& {Fall}, S.~M. 2000, \apj, 539, 718

\bibitem[{{Cielo} {et~al.}(2018){Cielo}, {Bieri}, {Volonteri}, {Wagner}, \&
  {Dubois}}]{Cielo2018}
{Cielo}, S., {Bieri}, R., {Volonteri}, M., {Wagner}, A.~Y., \& {Dubois}, Y.
  2018, \mnras, 477, 1336

\bibitem[{{Ciotti} {et~al.}(2010){Ciotti}, {Ostriker}, \& {Proga}}]{Ciotti2010}
{Ciotti}, L., {Ostriker}, J.~P., \& {Proga}, D. 2010, \apj, 717, 708

\bibitem[{{da Cunha} {et~al.}(2008){da Cunha}, {Charlot}, \&
  {Elbaz}}]{daCunha2008}
{da Cunha}, E., {Charlot}, S., \& {Elbaz}, D. 2008, \mnras, 388, 1595

\bibitem[{{Draine} {et~al.}(2014){Draine}, {Aniano}, {Krause}, {Groves},
  {Sandstrom}, {Braun}, {Leroy}, {Klaas}, {Linz}, {Rix}, {Schinnerer},
  {Schmiedeke}, \& {Walter}}]{dl2014}
{Draine}, B.~T., {Aniano}, G., {Krause}, O., {et~al.} 2014, \apj, 780, 172

\bibitem[{{Esposito} {et~al.}(2022){Esposito}, {Vallini}, {Pozzi}, {Casasola},
  {Mingozzi}, {Vignali}, {Gruppioni}, \& {Salvestrini}}]{Esposito22}
{Esposito}, F., {Vallini}, L., {Pozzi}, F., {et~al.} 2022, \mnras, 512, 686

\bibitem[{{Ferland} {et~al.}(2017){Ferland}, {Chatzikos}, {Guzm{\'a}n},
  {Lykins}, {van Hoof}, {Williams}, {Abel}, {Badnell}, {Keenan}, {Porter}, \&
  {Stancil}}]{Ferland2017}
{Ferland}, G.~J., {Chatzikos}, M., {Guzm{\'a}n}, F., {et~al.} 2017, \rmxaa, 53,
  385

\bibitem[{{Fierlinger} {et~al.}(2016){Fierlinger}, {Burkert}, {Ntormousi},
  {Fierlinger}, {Schartmann}, {Ballone}, {Krause}, \& {Diehl}}]{Fierlinger2016}
{Fierlinger}, K.~M., {Burkert}, A., {Ntormousi}, E., {et~al.} 2016, \mnras,
  456, 710

\bibitem[{{Flower} \& {Pineau des For{\^e}ts}(1999)}]{Flower1999}
{Flower}, D.~R. \& {Pineau des For{\^e}ts}, G. 1999, \mnras, 308, 271

\bibitem[{{Flower} \& {Pineau des For{\^e}ts}(2003)}]{FlowerPdF03}
{Flower}, D.~R. \& {Pineau des For{\^e}ts}, G. 2003, \mnras, 343, 390

\bibitem[{{Godard} {et~al.}(2019){Godard}, {Pineau des For{\^e}ts}, {Lesaffre},
  {Lehmann}, {Gusdorf}, \& {Falgarone}}]{Godard19}
{Godard}, B., {Pineau des For{\^e}ts}, G., {Lesaffre}, P., {et~al.} 2019, \aap,
  622, A100

\bibitem[{{Grieco} {et~al.}(2023){Grieco}, {Theul{\'e}}, {De Looze}, \&
  {Dulieu}}]{Grieco23}
{Grieco}, F., {Theul{\'e}}, P., {De Looze}, I., \& {Dulieu}, F. 2023, Nature
  Astronomy, 7, 541

\bibitem[{{Guillard} {et~al.}(2015{\natexlab{a}}){Guillard}, {Boulanger},
  {Lehnert}, {Appleton}, \& {Pineau des For{\^e}ts}}]{Guillard15a}
{Guillard}, P., {Boulanger}, F., {Lehnert}, M.~D., {Appleton}, P.~N., \&
  {Pineau des For{\^e}ts}, G. 2015{\natexlab{a}}, in SF2A-2015: Proceedings of
  the Annual meeting of the French Society of Astronomy and Astrophysics,
  81--85

\bibitem[{{Guillard} {et~al.}(2015{\natexlab{b}}){Guillard}, {Boulanger},
  {Lehnert}, {Pineau des For{\^e}ts}, {Combes}, {Falgarone}, \&
  {Bernard-Salas}}]{Guillard15}
{Guillard}, P., {Boulanger}, F., {Lehnert}, M.~D., {et~al.} 2015{\natexlab{b}},
  \aap, 574, A32

\bibitem[{{Guillard} {et~al.}(2009){Guillard}, {Boulanger}, {Pineau des
  For{\^e}ts}, \& {Appleton}}]{Guillard2009}
{Guillard}, P., {Boulanger}, F., {Pineau des For{\^e}ts}, G., \& {Appleton},
  P.~N. 2009, \aap, 502, 515

\bibitem[{{Guillard} {et~al.}(2012){Guillard}, {Ogle}, {Emonts}, {Appleton},
  {Morganti}, {Tadhunter}, {Oosterloo}, {Evans}, \& {Evans}}]{Guillard12}
{Guillard}, P., {Ogle}, P.~M., {Emonts}, B.~H.~C., {et~al.} 2012, \apj, 747, 95

\bibitem[{{Habart} {et~al.}(2005){Habart}, {Walmsley}, {Verstraete}, {Cazaux},
  {Maiolino}, {Cox}, {Boulanger}, \& {Pineau des For{\^e}ts}}]{Habart2005}
{Habart}, E., {Walmsley}, M., {Verstraete}, L., {et~al.} 2005, \ssr, 119, 71

\bibitem[{{Hardcastle} \& {Croston}(2020)}]{Hardcastle20}
{Hardcastle}, M.~J. \& {Croston}, J.~H. 2020, \nar, 88, 101539

\bibitem[{{Herrera} {et~al.}(2012){Herrera}, {Boulanger}, {Nesvadba}, \&
  {Falgarone}}]{Herrera12}
{Herrera}, C.~N., {Boulanger}, F., {Nesvadba}, N.~P.~H., \& {Falgarone}, E.
  2012, \aap, 538, L9

\bibitem[{Kim {et~al.}(2022)Kim, Gong, Kim, \& Ostriker}]{Kim2023}
Kim, J.-G., Gong, M., Kim, C.-G., \& Ostriker, E.~C. 2022, The Astrophysical
  Journal Supplement Series, 264, 10

\bibitem[{{Klessen} \& {Hennebelle}(2010)}]{Klessen2010}
{Klessen}, R.~S. \& {Hennebelle}, P. 2010, \aap, 520, A17

\bibitem[{{Krause}(2023)}]{Krause23}
{Krause}, M. G.~H. 2023, Galaxies, 11, 29

\bibitem[{{Kristensen} {et~al.}(2023){Kristensen}, {Godard}, {Guillard},
  {Gusdorf}, \& {Pineau des For{\^e}ts}}]{Kristensen23}
{Kristensen}, L.~E., {Godard}, B., {Guillard}, P., {Gusdorf}, A., \& {Pineau
  des For{\^e}ts}, G. 2023, \aap, 675, A86

\bibitem[{{Krumholz} {et~al.}(2014){Krumholz}, {Bate}, {Arce}, {Dale},
  {Gutermuth}, {Klein}, {Li}, {Nakamura}, \& {Zhang}}]{Krumholz2014}
{Krumholz}, M.~R., {Bate}, M.~R., {Arce}, H.~G., {et~al.} 2014, in Protostars
  and Planets VI, ed. H.~{Beuther}, R.~S. {Klessen}, C.~P. {Dullemond}, \&
  T.~{Henning}, 243--266

\bibitem[{{Lanz} {et~al.}(2016){Lanz}, {Ogle}, {Alatalo}, \&
  {Appleton}}]{Lanz2016}
{Lanz}, L., {Ogle}, P.~M., {Alatalo}, K., \& {Appleton}, P.~N. 2016, \apj, 826,
  29

\bibitem[{{Le Petit} {et~al.}(2006){Le Petit}, {Nehm{\'e}}, {Le Bourlot}, \&
  {Roueff}}]{LePetit2006}
{Le Petit}, F., {Nehm{\'e}}, C., {Le Bourlot}, J., \& {Roueff}, E. 2006, \apjs,
  164, 506

\bibitem[{{Leftley} {et~al.}(2024){Leftley}, {Nesvadba}, {Bicknell}, {Janssen},
  {Mukherjee}, {Petrov}, {Shende}, \& {Zovaro}}]{Leftley24}
{Leftley}, J.~H., {Nesvadba}, N. P.~H., {Bicknell}, G., {et~al.} 2024, arXiv
  e-prints, arXiv:2404.04341

\bibitem[{{Lehmann} {et~al.}(2016){Lehmann}, {Federrath}, \&
  {Wardle}}]{Lehamnn16}
{Lehmann}, A., {Federrath}, C., \& {Wardle}, M. 2016, \mnras, 463, 1026

\bibitem[{{Lehmann} {et~al.}(2022){Lehmann}, {Godard}, {Pineau des For{\^e}ts},
  {Vidal-Garc{\'\i}a}, \& {Falgarone}}]{Lehmann22}
{Lehmann}, A., {Godard}, B., {Pineau des For{\^e}ts}, G., {Vidal-Garc{\'\i}a},
  A., \& {Falgarone}, E. 2022, \aap, 658, A165

\bibitem[{{Lesaffre} {et~al.}(2013){Lesaffre}, {Pineau des For{\^e}ts},
  {Godard}, {Guillard}, {Boulanger}, \& {Falgarone}}]{Lesaffre2013}
{Lesaffre}, P., {Pineau des For{\^e}ts}, G., {Godard}, B., {et~al.} 2013, \aap,
  550, A106

\bibitem[{{Lesaffre} {et~al.}(2020){Lesaffre}, {Todorov}, {Levrier},
  {Valdivia}, {Dzyurkevich}, {Godard}, {Tram}, {Gusdorf}, {Lehmann}, \&
  {Falgarone}}]{Lesaffre2020}
{Lesaffre}, P., {Todorov}, P., {Levrier}, F., {et~al.} 2020, \mnras, 495, 816

\bibitem[{{Lilly} \& {Longair}(1984)}]{Lilly84}
{Lilly}, S.~J. \& {Longair}, M.~S. 1984, \mnras, 211, 833

\bibitem[{{Mathis} {et~al.}(1983){Mathis}, {Mezger}, \& {Panagia}}]{Mathis83}
{Mathis}, J.~S., {Mezger}, P.~G., \& {Panagia}, N. 1983, \aap, 128, 212

\bibitem[{{Meijerink} {et~al.}(2013){Meijerink}, {Kristensen}, {Wei{\ss}}, {van
  der Werf}, {Walter}, {Spaans}, {Loenen}, {Fischer}, {Israel}, {Isaak},
  {Papadopoulos}, {Aalto}, {Armus}, {Charmandaris}, {Dasyra}, {Diaz-Santos},
  {Evans}, {Gao}, {Gonz{\'a}lez-Alfonso}, {G{\"u}sten}, {Henkel}, {Kramer},
  {Lord}, {Mart{\'\i}n-Pintado}, {Naylor}, {Sanders}, {Smith}, {Spinoglio},
  {Stacey}, {Veilleux}, \& {Wiedner}}]{Meijerink13}
{Meijerink}, R., {Kristensen}, L.~E., {Wei{\ss}}, A., {et~al.} 2013, \apjl,
  762, L16

\bibitem[{{Mingozzi} {et~al.}(2018){Mingozzi}, {Vallini}, {Pozzi}, {Vignali},
  {Mignano}, {Gruppioni}, {Talia}, {Cimatti}, {Cresci}, \&
  {Massardi}}]{Mingozzi18}
{Mingozzi}, M., {Vallini}, L., {Pozzi}, F., {et~al.} 2018, \mnras, 474, 3640

\bibitem[{{Morganti} {et~al.}(2023){Morganti}, {Murthy}, {Guillard},
  {Oosterloo}, \& {Garcia-Burillo}}]{Morganti23}
{Morganti}, R., {Murthy}, S., {Guillard}, P., {Oosterloo}, T., \&
  {Garcia-Burillo}, S. 2023, Galaxies, 11, 24

\bibitem[{{Mukherjee} {et~al.}(2016){Mukherjee}, {Bicknell}, {Sutherland}, \&
  {Wagner}}]{Mukherjee16}
{Mukherjee}, D., {Bicknell}, G.~V., {Sutherland}, R., \& {Wagner}, A. 2016,
  \mnras, 461, 967

\bibitem[{{Mukherjee} {et~al.}(2018){Mukherjee}, {Bicknell}, {Wagner},
  {Sutherland}, \& {Silk}}]{Mukherjee18}
{Mukherjee}, D., {Bicknell}, G.~V., {Wagner}, A.~Y., {Sutherland}, R.~S., \&
  {Silk}, J. 2018, \mnras, 479, 5544

\bibitem[{{Murthy} {et~al.}(2022){Murthy}, {Morganti}, {Wagner}, {Oosterloo},
  {Guillard}, {Mukherjee}, \& {Bicknell}}]{Murthy22}
{Murthy}, S., {Morganti}, R., {Wagner}, A.~Y., {et~al.} 2022, Nature Astronomy,
  6, 488

\bibitem[{{Nesvadba} {et~al.}(2011){Nesvadba}, {Boulanger}, {Lehnert},
  {Guillard}, \& {Salome}}]{Nesvadba11}
{Nesvadba}, N.~P.~H., {Boulanger}, F., {Lehnert}, M.~D., {Guillard}, P., \&
  {Salome}, P. 2011, \aap, 536, L5

\bibitem[{{Nesvadba} {et~al.}(2010){Nesvadba}, {Boulanger}, {Salom{\'e}},
  {Guillard}, {Lehnert}, {Ogle}, {Appleton}, {Falgarone}, \& {Pineau Des
  Forets}}]{Nesvadba10}
{Nesvadba}, N.~P.~H., {Boulanger}, F., {Salom{\'e}}, P., {et~al.} 2010, \aap,
  521, A65

\bibitem[{{Ogle} {et~al.}(2007){Ogle}, {Antonucci}, {Appleton}, \&
  {Whysong}}]{Ogle07}
{Ogle}, P., {Antonucci}, R., {Appleton}, P.~N., \& {Whysong}, D. 2007, \apj,
  668, 699

\bibitem[{{Ogle} {et~al.}(2010){Ogle}, {Boulanger}, {Guillard}, {Evans},
  {Antonucci}, {Appleton}, {Nesvadba}, \& {Leipski}}]{Ogle10}
{Ogle}, P., {Boulanger}, F., {Guillard}, P., {et~al.} 2010, \apj, 724, 1193

\bibitem[{{Ogle} {et~al.}(2024){Ogle}, {L{\'o}pez}, {Reynaldi}, {Togi}, {Rich},
  {Rom{\'a}n}, {Caceres}, {Li}, {Donnelly}, {Smith}, {Appleton}, \&
  {Lanz}}]{Ogle23}
{Ogle}, P.~M., {L{\'o}pez}, I.~E., {Reynaldi}, V., {et~al.} 2024, \apj, 962,
  196

\bibitem[{{Ostriker} \& {Kim}(2022)}]{Ostriker2022}
{Ostriker}, E.~C. \& {Kim}, C.-G. 2022, \apj, 936, 137

\bibitem[{{Park} \& {Ryu}(2019)}]{Park2019}
{Park}, J. \& {Ryu}, D. 2019, \apj, 875, 2

\bibitem[{{Pathak} {et~al.}(2024){Pathak}, {Leroy}, {Thompson}, {Lopez},
  {Belfiore}, {Boquien}, {Dale}, {Glover}, {Klessen}, {Koch}, {Rosolowsky},
  {Sandstrom}, {Schinnerer}, {Smith}, {Sun}, {Sutter}, {Williams}, {Bigiel},
  {Cao}, {Chastenet}, {Chevance}, {Chown}, {Emsellem}, {Faesi}, {Larson},
  {Lee}, {Meidt}, {Ostriker}, {Ramambason}, {Sarbadhicary}, \&
  {Thilker}}]{Pathak2024}
{Pathak}, D., {Leroy}, A.~K., {Thompson}, T.~A., {et~al.} 2024, \aj, 167, 39

\bibitem[{{Pereira-Santaella} {et~al.}(2022){Pereira-Santaella},
  {{\'A}lvarez-M{\'a}rquez}, {Garc{\'\i}a-Bernete}, {Labiano}, {Colina},
  {Alonso-Herrero}, {Bellocchi}, {Garc{\'\i}a-Burillo}, {H{\"o}nig}, {Ramos
  Almeida}, \& {Rosario}}]{Pereira-Santaella2022}
{Pereira-Santaella}, M., {{\'A}lvarez-M{\'a}rquez}, J., {Garc{\'\i}a-Bernete},
  I., {et~al.} 2022, \aap, 665, L11

\bibitem[{{Pfenniger}(2010)}]{Pfenniger2010}
{Pfenniger}, D. 2010, in Astronomical Society of the Pacific Conference Series,
  Vol. 421, Galaxies in Isolation: Exploring Nature Versus Nurture, ed.
  L.~{Verdes-Montenegro}, A.~{Del Olmo}, \& J.~{Sulentic}, 183

\bibitem[{{Planck Collaboration} {et~al.}(2020){Planck Collaboration},
  {Aghanim}, {Akrami}, {Ashdown}, {Aumont}, {Baccigalupi}, {Ballardini},
  {Banday}, {Barreiro}, {Bartolo}, {Basak}, {Battye}, {Benabed}, {Bernard},
  {Bersanelli}, {Bielewicz}, {Bock}, {Bond}, {Borrill}, {Bouchet}, {Boulanger},
  {Bucher}, {Burigana}, {Butler}, {Calabrese}, {Cardoso}, {Carron},
  {Challinor}, {Chiang}, {Chluba}, {Colombo}, {Combet}, {Contreras}, {Crill},
  {Cuttaia}, {de Bernardis}, {de Zotti}, {Delabrouille}, {Delouis}, {Di
  Valentino}, {Diego}, {Dor{\'e}}, {Douspis}, {Ducout}, {Dupac}, {Dusini},
  {Efstathiou}, {Elsner}, {En{\ss}lin}, {Eriksen}, {Fantaye}, {Farhang},
  {Fergusson}, {Fernandez-Cobos}, {Finelli}, {Forastieri}, {Frailis},
  {Fraisse}, {Franceschi}, {Frolov}, {Galeotta}, {Galli}, {Ganga},
  {G{\'e}nova-Santos}, {Gerbino}, {Ghosh}, {Gonz{\'a}lez-Nuevo}, {G{\'o}rski},
  {Gratton}, {Gruppuso}, {Gudmundsson}, {Hamann}, {Handley}, {Hansen},
  {Herranz}, {Hildebrandt}, {Hivon}, {Huang}, {Jaffe}, {Jones}, {Karakci},
  {Keih{\"a}nen}, {Keskitalo}, {Kiiveri}, {Kim}, {Kisner}, {Knox},
  {Krachmalnicoff}, {Kunz}, {Kurki-Suonio}, {Lagache}, {Lamarre}, {Lasenby},
  {Lattanzi}, {Lawrence}, {Le Jeune}, {Lemos}, {Lesgourgues}, {Levrier},
  {Lewis}, {Liguori}, {Lilje}, {Lilley}, {Lindholm}, {L{\'o}pez-Caniego},
  {Lubin}, {Ma}, {Mac{\'\i}as-P{\'e}rez}, {Maggio}, {Maino}, {Mandolesi},
  {Mangilli}, {Marcos-Caballero}, {Maris}, {Martin}, {Martinelli},
  {Mart{\'\i}nez-Gonz{\'a}lez}, {Matarrese}, {Mauri}, {McEwen}, {Meinhold},
  {Melchiorri}, {Mennella}, {Migliaccio}, {Millea}, {Mitra},
  {Miville-Desch{\^e}nes}, {Molinari}, {Montier}, {Morgante}, {Moss}, {Natoli},
  {N{\o}rgaard-Nielsen}, {Pagano}, {Paoletti}, {Partridge}, {Patanchon},
  {Peiris}, {Perrotta}, {Pettorino}, {Piacentini}, {Polastri}, {Polenta},
  {Puget}, {Rachen}, {Reinecke}, {Remazeilles}, {Renzi}, {Rocha}, {Rosset},
  {Roudier}, {Rubi{\~n}o-Mart{\'\i}n}, {Ruiz-Granados}, {Salvati}, {Sandri},
  {Savelainen}, {Scott}, {Shellard}, {Sirignano}, {Sirri}, {Spencer},
  {Sunyaev}, {Suur-Uski}, {Tauber}, {Tavagnacco}, {Tenti}, {Toffolatti},
  {Tomasi}, {Trombetti}, {Valenziano}, {Valiviita}, {Van Tent}, {Vibert},
  {Vielva}, {Villa}, {Vittorio}, {Wandelt}, {Wehus}, {White}, {White},
  {Zacchei}, \& {Zonca}}]{Planck2020}
{Planck Collaboration}, {Aghanim}, N., {Akrami}, Y., {et~al.} 2020, \aap, 641,
  A6

\bibitem[{{Polles} {et~al.}(2021){Polles}, {Salom{\'e}}, {Guillard}, {Godard},
  {Pineau des For{\^e}ts}, {Olivares}, {Beckmann}, {Canning}, {Combes},
  {Dubois}, {Edge}, {Fabian}, {Ferland}, {Hamer}, \& {Lehnert}}]{Polles2021}
{Polles}, F.~L., {Salom{\'e}}, P., {Guillard}, P., {et~al.} 2021, \aap, 651,
  A13

\bibitem[{{Richard} {et~al.}(2022){Richard}, {Lesaffre}, {Falgarone}, \&
  {Lehmann}}]{Richard22}
{Richard}, T., {Lesaffre}, P., {Falgarone}, E., \& {Lehmann}, A. 2022, \aap,
  664, A193

\bibitem[{{Shaw} {et~al.}(2009){Shaw}, {Ferland}, {Henney}, {Stancil}, {Abel},
  {Pellegrini}, {Baldwin}, \& {van Hoof}}]{Shaw2009}
{Shaw}, G., {Ferland}, G.~J., {Henney}, W.~J., {et~al.} 2009, \apj, 701, 677

\bibitem[{{Stalevski} {et~al.}(2016){Stalevski}, {Ricci}, {Ueda}, {Lira},
  {Fritz}, \& {Baes}}]{Stalevski2016}
{Stalevski}, M., {Ricci}, C., {Ueda}, Y., {et~al.} 2016, \mnras, 458, 2288

\bibitem[{{Vazquez-Semadeni}(2009)}]{Vazquez-Semadeni2009}
{Vazquez-Semadeni}, E. 2009, arXiv e-prints, arXiv:0902.0820

\bibitem[{{Willis} \& {Strom}(1978)}]{Willis78}
{Willis}, A.~G. \& {Strom}, R.~G. 1978, \aap, 62, 375

\bibitem[{{Wittor} \& {Gaspari}(2020)}]{Wittor2020}
{Wittor}, D. \& {Gaspari}, M. 2020, \mnras, 498, 4983

\bibitem[{{Yang} {et~al.}(2022){Yang}, {Boquien}, {Brandt}, {Buat},
  {Burgarella}, {Ciesla}, {Lehmer}, {Ma{\l}ek}, {Mountrichas}, {Papovich},
  {Pons}, {Stalevski}, {Theul{\'e}}, \& {Zhu}}]{Yang22}
{Yang}, G., {Boquien}, M., {Brandt}, W.~N., {et~al.} 2022, \apj, 927, 192

\bibitem[{{Yang} {et~al.}(2020){Yang}, {Boquien}, {Buat}, {Burgarella},
  {Ciesla}, {Duras}, {Stalevski}, {Brandt}, \& {Papovich}}]{Yang20}
{Yang}, G., {Boquien}, M., {Buat}, V., {et~al.} 2020, \mnras, 491, 740

\end{thebibliography}

\begin{appendix}


\section{Conditions of validity of the toy model}\label{sec:Lehman_toymodel}

\begin{figure*}
    \centering
     \includegraphics[width=0.8\textwidth,clip]{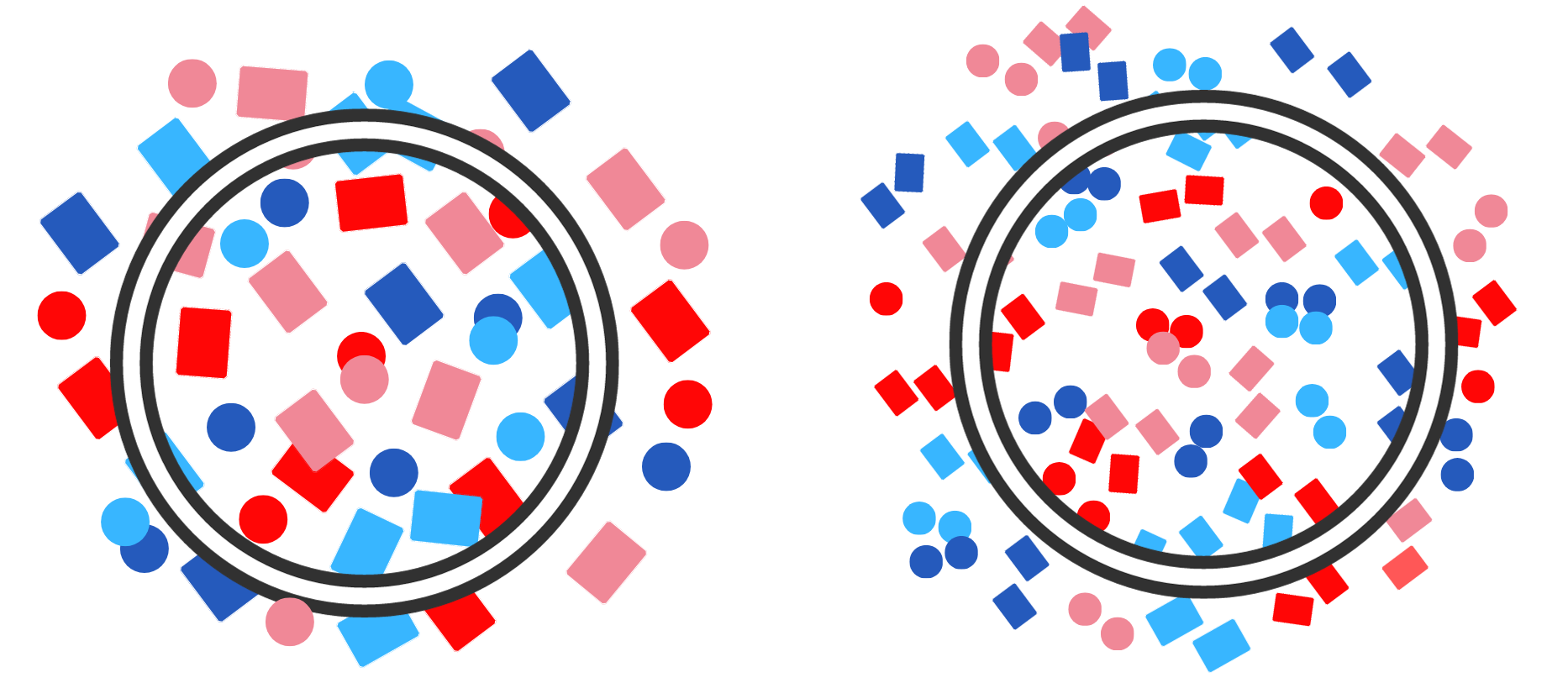}
    \caption{Schematic representation of ensembles of shocks and PDRs inside an observational beam with a solid angle $\Omega_{\rm obs}$ (black and white circle). Shock or PDR surfaces are represented by the filled circles and filled squares, respectively, and are randomly placed within the beam. The color represents the systemic velocities of shock and PDR surfaces projected along the line of sight. Blue (resp. red) circles and squares correspond to blueshifted (resp. redshifted) shocks and PDRs with respect to the recession velocity of the galaxy. 
    The \textit{left} and \textit{right} panels represent two equivalent scenarios, where the numbers of shock (resp. PDR) surfaces are different, but subtend the same total solid $\Omega_{\rm S}$ (resp. $\Omega_{\rm P}$). This figure is adapted from \cite{Lehmann22}.}
    \label{fig1_2}
\end{figure*}

\begin{figure*}
    \centering
    \includegraphics[width=0.81\textwidth,clip]{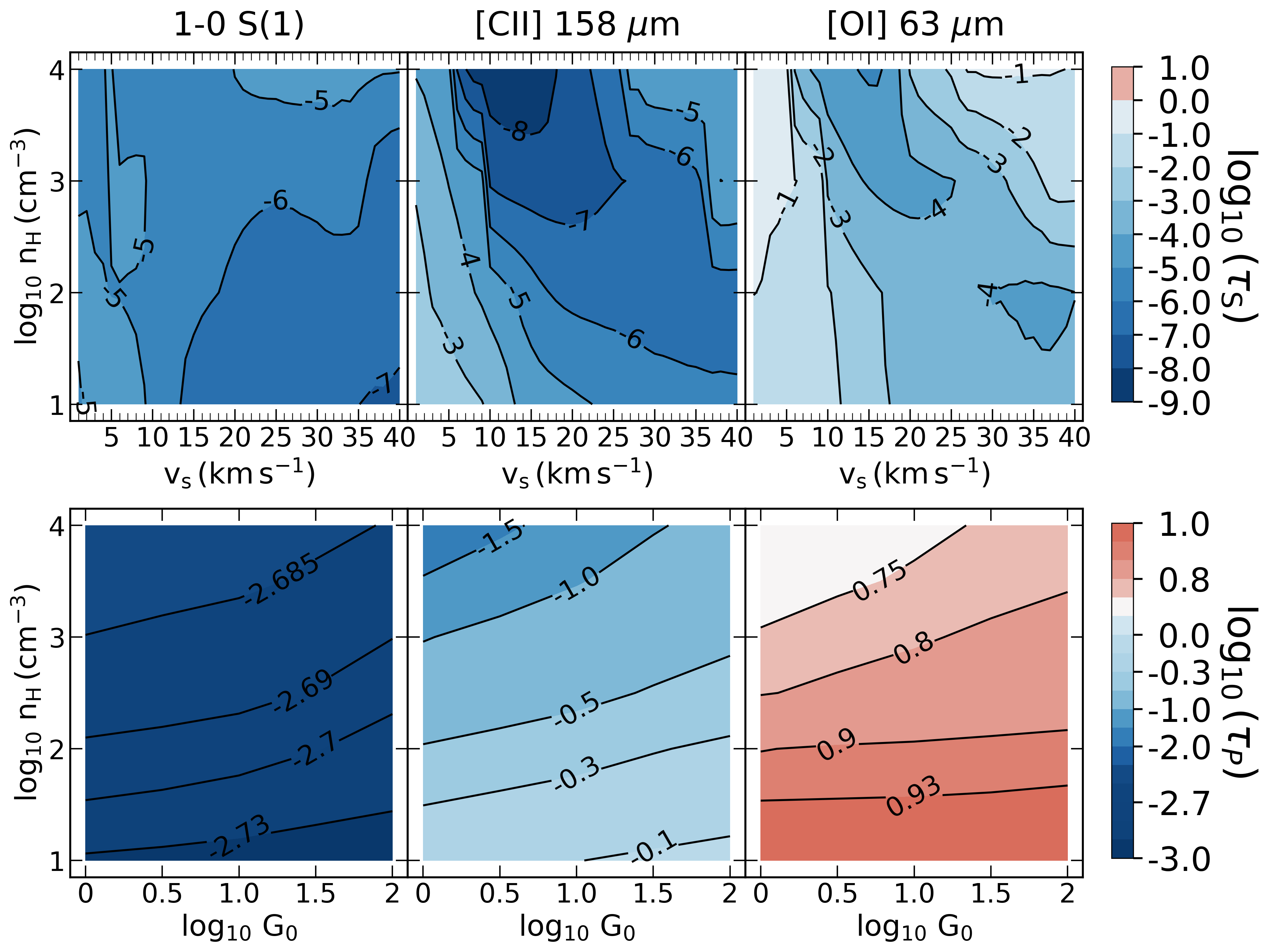} \\
    \caption{Opacities of the emission lines in shocks as functions of the pre-shock density and the shock velocity (top row) and in PDRs as function of the gas density and the illumination factor (bottom row). All other parameters are set to their standard values given in Table~\ref{table:grid_params}. The left column is associated with the H$_2$ 1-0 S(1) line, the middle column with the [CII]~$158~\mu$m line, and the right column to [OI]~$63~\mu$m line. The color code indicates the central opacity of the lines $\tau_{\rm S}(\nu_0)$ and $\tau_{\rm P}(\nu_0)$, with a different color scale for shocks and PDRs.}
    \label{fig_tau}
\end{figure*}

The toy model presented in Sect. \ref{sec:distribution-shocks} and used throughout this paper results from a simplification of the radiative transfer of an ensemble of shocks and PDRs within an observational beam. As illustrated in Fig. \ref{fig1_2}, we consider that an observational beam of solid angle $\Omega_{\rm obs}$ contains shock and PDR surfaces with a distribution of systemic velocities and with total solid angles $\Omega_{\rm S}$ and $\Omega_{\rm P}$. For simplicity, we set that the emission from one shock or one PDR in a line with a rest frequency $\nu_0$ follows a tophat profile over a width $\Delta \nu = \nu_0 \Delta V / c$, with a velocity width $\Delta V = 5$ km s$^{-1}$ dominated by turbulent motions \citep{Lehmann22}. Both shock and PDR surfaces are assumed to have systemic velocities projected along the observed line of sight that follow a Gaussian distribution in frequency space
\begin{equation}
\phi_\nu = \frac{1}{\sqrt{2\pi}\sigma_\nu^{\rm obs}} {\rm exp}\left( - \frac{(\nu-\nu_0)^2}{2 \sigma_\nu^{{\rm obs}\,2}}\right),
\end{equation}
where $\sigma_\nu^{\rm obs}$ is the observed frequency dispersion of the line. Independently of this distribution, we also assume that $\Omega_{\rm S}$ and $\Omega_{\rm P}$ follow 1D probability distribution functions, $f_{\rm S}(\Theta_{\rm S})$ and $f_{\rm P}(\Theta_{\rm P})$, which both depend on a unique parameter, $\Theta_{\rm S}$ for the shocks and $\Theta_{\rm P}$ for the PDRs\footnote{This approach can easily be expanded to multidimensional distributions. In this paper, we assume a distribution of shock velocities, that is $\Theta_{\rm S} = {\rm v_S}$, and a distribution of illumination factors for the PDRs, that is $\Theta_{\rm P} = G_0$.}. The solid angles of shock and PDR surfaces that emit within a frequency bin of size $\Delta \nu$ centered at a frequency $\nu$ therefore write
\begin{equation}
\Delta \Omega_{\rm S}(\nu) = \Omega_{\rm S} \phi_\nu \Delta \nu = \int_{\Theta_{\rm S}} f_{\rm S}(\Theta_{\rm S}) d \Theta_{\rm S} \phi_\nu \Delta \nu
\end{equation}
and
\begin{equation}
\Delta \Omega_{\rm P}(\nu) = \Omega_{\rm P} \phi_\nu \Delta \nu = \int_{\Theta_{\rm P}} f_{\rm P}(\Theta_{\rm P}) d \Theta_{\rm P} \phi_\nu \Delta \nu.
\end{equation}
The number of shock and PDR layers within a frequency bin or, equivalently, the beam-filling factors per frequency bin of shocks and PDRs are
\begin{equation}
\frac{\Delta \Omega_{\rm S}}{\Omega_{\rm obs}}(\nu)
\end{equation}
and
\begin{equation}
\frac{\Delta \Omega_{\rm P}}{\Omega_{\rm obs}}(\nu).
\end{equation}
Their average total opacities are
\begin{equation}
\overline{\tau}_{\rm S}(\nu) = \int_{\Theta_{\rm S}} f_{\rm S}(\Theta_{\rm S}) \tau_{\rm S}(\Theta_{\rm S}) d \Theta_{\rm S} \phi_\nu \Delta \nu / \Omega_{\rm obs}
\end{equation}
and
\begin{equation}
\overline{\tau}_{\rm P}(\nu) = \int_{\Theta_{\rm P}} f_{\rm P}(\Theta_{\rm P}) \tau_{\rm P}(\Theta_{\rm P}) d \Theta_{\rm P} \phi_\nu \Delta \nu / \Omega_{\rm obs},
\end{equation}
where $\tau_{\rm S}(\Theta_{\rm S})$ and $\tau_{\rm P}(\Theta_{\rm P})$ are the opacities across one shock layer of parameter $\Theta_{\rm S}$ and one PDR layer of parameter $\Theta_{\rm P}$ respectively (see Fig. \ref{fig_tau}).

If there are no spatial overlaps of shocks and PDRs within any frequency bin, that is if
\begin{equation} \label{Eq-cond1}
\frac{\Delta \Omega_{\rm S}}{\Omega_{\rm obs}}(\nu_0) \ll 1 \quad {\rm and} \quad \frac{\Delta \Omega_{\rm P}}{\Omega_{\rm obs}}(\nu_0) \ll 1,
\end{equation} 
or if the averaged line opacities across the shock and PDR layers in any frequency bin are small, 
\begin{equation} \label{Eq-cond2}
\overline{\tau}_{\rm S}(\nu_0) \ll 1  \quad {\rm and} \quad \overline{\tau}_{\rm P}(\nu_0) \ll 1,
\end{equation}
absorption by shocks and PDRs can be neglected. The density flux (in erg cm$^{-2}$ s$^{-1}$ Hz$^{-1}$) collected in the observational beam therefore simply writes
\begin{equation} 
F_\nu = I_\nu^c \Omega_{\rm obs} 
+ \int_{\Theta_{\rm S}} f_{\rm S}(\Theta_{\rm S}) \phi_\nu I_{\rm S}(\Theta_{\rm S}) d\Theta_{\rm S}
+ \int_{\Theta_{\rm P}} f_{\rm P}(\Theta_{\rm P}) \phi_\nu I_{\rm P}(\Theta_{\rm P}) d\Theta_{\rm S}
\end{equation}
where $I_\nu^c$ is the specific intensity of the background continuum (in erg s$^{-1}$ cm$^{-2}$ Hz$^{-1}$ sr$^{-1}$) and $I_{\rm S}(\Theta_{\rm S})$ and $I_{\rm P}(\Theta_{\rm P})$ are the line integrated intensities emitted by one shock of parameter $\Theta_{\rm S}$ and one PDR of parameter $\Theta_{\rm P}$ respectively (in erg s$^{-1}$ cm$^{-2}$ sr$^{-1}$). Integrating this equation over frequency finally gives the continuum subtracted line flux (in erg s$^{-1}$ cm$^{-2}$)
\begin{equation} \label{Eq-final}
F = \int_{\Theta_{\rm S}} f_{\rm S}(\Theta_{\rm S}) I_{\rm S}(\Theta_{\rm S}) d \Theta_{\rm S} + \int_{\Theta_{\rm P}} f_{\rm P}(\Theta_{\rm P}) I_{\rm P}(\Theta_{\rm P}) d \Theta_{\rm P}.
\end{equation}
This last equation is at the root of the toy model presented in Sect. \ref{sec:distribution-shocks} (Eq. \ref{Eq-Imod}). It results from the assumption that the absorption induced by shock and PDR surfaces within the observational beam is negligible. Throughout the paper, we consider that Eq. \ref{Eq-final} is always valid and can be used to find the optimal combination of models that reproduces a given set of observations (see Sect. \ref{sec:results}). The validity of this assumption is checked a posteriori by verifying the conditions written in Eqs. \ref{Eq-cond1} and \ref{Eq-cond2}. The domain of parameters where these conditions break down are highlighted in Figs. \ref{fig0_2} and \ref{fig0_2_app}.

\section{The radio galaxy 3C~326~N}\label{sec:3C326source}


3C 326 is a well-known radio source located at $\rm z \sim 0.089\pm0.001$ \citep{Ogle07}. Optical and IR imaging shows that it is a pair of interacting galaxies, 3C~326~N and 3C 326 S (Fig.~\ref{fig1_1}). We focus on 3C~326~N, centered at $\alpha$(J2000) = $\rm 15^h 52^m 9^s.178$, $\delta$(J2000) = $\rm + 20^{deg} 05^m 48^s.23$, which hosts the AGN powering the radio jet observed on large scales, with a kinetic luminosity $\rm L_{jet} \sim 10^{11}~L_\odot$ \citep{Nesvadba10}. This galaxy has a remarkably low star-formation rate $\rm SFR = (7\pm3)\times10^{-2}~M_\odot~yr^{-1}$  \citep{Ogle07}, yet a large molecular gas content $\rm M_{gas} \sim 2\times10^{9}$~\Msun~\citep{Guillard15}. Strong IR H$_2$ line emission has been detected, with a total luminosity ${\rm L(H_2)} \approx 2.4 \times  10^8$~\Lsun~\citep{Ogle07}.  3C~326~N represents a perfect target to test which type of physical mechanisms (e.g., shocks and/or PDRs) are responsible for powering the ensemble of H$_2$ lines while putting constraints on the energy budget of the galaxy. Based on UV GALEX measurements, \cite{Ogle10} derived a value of G$_0 = 6^{+4}_{-2}$ for the radiation field in 3C~326~N. Similarly, \cite{Guillard15} reported a value of G$_0 = 9 \pm 1$ from a fit of the SED in the IR range.

Table \ref{table:0_1} provides a summary of the relevant physical parameters of 3C~326~N\footnote{With the adopted cosmology, the 3C~326 redshift of $z = 0.089$ corresponds to a luminosity distance D$_{\rm L}$ = 422 Mpc, an angular diameter distance D$_{\rm A}$ = 356 Mpc, and a scale of 1.725 kpc/arcsec.}. The observational solid angle is estimated from the source diameter in optical and IRAC images \citep{Nesvadba10}. The bulk of the H$_2$ emission is contained inside this beam \citep{Ogle10, Nesvadba11} as well as the [CII]~$158~\mu$m and [OI]~$63~\mu$m emissions \citep{Guillard15}. The observed fluxes given in Table \ref{table:1} are corrected for redshift as
\begin{equation} \label{Eq-redshift}
{F}^{\rm corr} = {F}^{\rm obs}\,\,(1 + \rm{z})^3
\end{equation}
and transformed into integrated intensities using the observational solid angle $\Omega_{\rm obs}$. We note that the total observed luminosity of $\rm H_2$, [CII]~$158~\mu$m and [OI]~$63~\mu$m lines is $\rm 2.5\times10^8~L_\odot$ which corresponds to about 3\% of the IR luminosity and 0.25\% of the jet’s luminosity.


\subsection{\texorpdfstring{H$_2$}{H2 pure rotational and ro-vibrational lines} pure rotational and ro-vibrational lines}\label{ssec:Shock_lines}

The observed fluxes and relevant information on the emission lines of molecular hydrogen are given in Table~\ref{table:1}. A total of twelve H$_2$ emission lines have been measured in 3C~326~N. Eight of these are the pure-rotational lines from H$_2$ 0-0 S(0) to H$_2$ 0-0 S(7), obtained with the mid-infrared \textit{Spitzer} IRS low-resolution spectroscopy ($\rm R\sim60-120$), which did not resolve the lines \citep{Ogle10}. Four ro-vibrational emission lines, from H$_2$ 1-0 S(1) to H$_2$ 1-0 S(4), were spatially and spectrally resolved by the VLT/SINFONI spectrograph 
\citep{Nesvadba11}. However, since the main goal is to fit the observed integrated intensity of each line, the spectral resolution does not have any impact on our analysis.

\subsection{[CII] \texorpdfstring{$158~\mu{\rm m}$}{}, [OI] \texorpdfstring{$63~\mu{\rm m}$}{} and CO(1-0)}\label{ssec:PDR_lines}

The observed fluxes and relevant information on the emission lines of C$^+$, O, and CO are given in Table~\ref{table:1}. The [CII]~$158~\mu$m and the [OI]~$63~\mu$m emission lines are taken from observations obtained with \textit{Herschel}/PACS and reported in \cite{Guillard15}.
The CO(1-0) line intensity is taken from observations performed with the IRAM PdBI and provided by \cite{Nesvadba10}. As explained in the main text, this CO(1-0) emission is not used to constrain the parameters of the model but is used as a sanity check to test the validity of the solutions obtained.

\begin{table}
\caption{Observational parameters of 3C~326~N.}     
\label{table:0_1} 
\centering 
\resizebox{\columnwidth}{!}{%
\begin{tabular}{lccc}
\hline\hline
 \textbf{Parameters} &  \textbf{Symbol} &  \textbf{Value} &  \textbf{Units}  \\
 \hline 
 $^{\rm a}$Observational solid angle & $\Omega_{\rm obs}$ & $13.69$ & arcsec$^2$  \\
 Angular diameter distance & D$_{\rm A}$ & $356$ & Mpc \\
 $^{\rm b}$Bolometric luminosity & L$_{\rm bol}$ & $\sim 2.6\times10^{10}$ & L$_\odot$  \\
 $^{\rm c}$IR luminosity & L$_{\rm IR}$ & $ (7.1\pm1.4)\times10^9$ & L$_\odot$  \\
  $^{\rm d}$Kinetic power available & L$_{\rm K}$ & $\sim 10^{9}$ & L$_\odot$  \\
 $^{\rm e}$Jet kinetic power & L$_{\rm jet}$ & $\sim 10^{11}$ & L$_\odot$  \\
 $^{\rm f}$Redshift & z & $0.089$ & ---  \\
 \hline
\end{tabular}

} \\

{\small
{\raggedright \textbf{Notes.} \par}
{\raggedright $^{\rm a}$ Solid angle derived from the source diameter in optical and IRAC images \citep{Nesvadba10}. \par}
{\raggedright $^{\rm b}$ Bolometric luminosity reported in \cite{Nesvadba10}.\par}
{\raggedright $^{\rm c}$ SED estimated value reported in Table  \ref{table_bayes_app}. \par}
{\raggedright $^{\rm d}$ Observational derived values \citep{Ogle07, Guillard15}. \par}
{\raggedright $^{\rm e}$ Estimated value in \cite{Nesvadba10}. \par}
{\raggedright $^{\rm f}$ Value reported in \cite{Ogle07}. \par}
}
\end{table}

\begin{table}

\caption{Emission line properties of 3C~326~N.}             
\label{table:1}      
\centering 

\resizebox{\columnwidth}{!}{%
\begin{tabular}{ccccc}

\hline\hline
 \textbf{Line} &  \textbf{$\lambda_{\rm rest}$} &  \textbf{Flux}  &  \textbf{FWHM}  & \textbf{SNR}  \\
               &  ($\mu{\rm m}$)                & ($\rm 10^{-15}~erg~s^{-1}~cm^{-2}$) & (km s$^{-1}$) &  \\
 \hline 


 0-0 S(0) & 28.22  & $3.0  \pm 0.6$ $^{\rm a}$ & --- & 5.0  \\
 0-0 S(1) & 17.03  & $6.9  \pm 0.6$ $^{\rm a}$ & --- & 11.5 \\
 0-0 S(2) & 12.28  & $4.1  \pm 0.4$ $^{\rm a}$ & --- & 10.2 \\
 0-0 S(3) & 9.66   & $12.6 \pm 0.5$ $^{\rm a}$ & --- & 25.2 \\
 0-0 S(4) & 8.03   & $3.1  \pm 0.6$ $^{\rm a}$ & --- & 5.2  \\
 0-0 S(5) & 6.91   & $4.6  \pm 2.2$ $^{\rm a}$ & --- & 2.1  \\
 0-0 S(6) & 6.11   & $2.5  \pm 0.9$ $^{\rm a}$ & --- & 2.8  \\
 0-0 S(7) & 5.55   & $2.9  \pm 0.9$ $^{\rm a}$ & --- & 3.2  \\
 
 1-0 S(1) & 2.122  & $1.8 \pm 0.1$  $^{\rm b}$ & $534 \pm 32$ & 13.8 \\
 1-0 S(2) & 2.034  & $0.7 \pm 0.1$  $^{\rm b}$ & $563 \pm 45$ & 14.0 \\
 1-0 S(3) & 1.958  & $2.2 \pm 0.2$  $^{\rm b}$ & $594 \pm 41$ & 11.0 \\
 1-0 S(4) & 1.892  & $0.8 \pm 0.1$  $^{\rm b}$ & $475 \pm 45$ & 16.0 \\
 
 [CII]    & 157.74 & $11.8 \pm 1.0$ $^{\rm c}$ & $372 \pm 12$ & 11.8 \\
 $[\rm OI]$     & 63.18  & $6.2 \pm 1.4$ $^{\rm c}$ & $352 \pm 31$ & 4.7 \\
 CO(1-0)  & 2600   & $0.0035 \pm 0.0007$ $^{\rm d}$ & $352 \pm 58$ & 5.0 \\
 \hline
\end{tabular}
}
{\small
{\raggedright \textbf{Notes.} \par}
{\raggedright $^{\rm a}$ \cite{Ogle10} from observations using \textit{Spitzer} IRS. \par}
{\raggedright $^{\rm b}$ \cite{Nesvadba11} from observations using VLT/SINFONI. \par}
{\raggedright $^{\rm c}$ \cite{Guillard15} from observations using \textit{Herschel}/PACS. \par}
{\raggedright $^{\rm d}$ \cite{Nesvadba10} from observations using IRAM PdBI. \par}
}
\end{table}

\section{Spectral energy distribution}\label{ssec:SED}

The SED of 3C~326~N is constrained using the photometric data spanning from the X-ray to the radio bands and available on the NASA/IPAC Extragalactic Database (NED). In this paper, we use the measurements of 14 photometric bands obtained, in particular, by \cite{Lilly84}, \cite{Ogle10}, and \cite{Guillard15}, and given in Table~\ref{table:phot}. These bands include the ultraviolet NUV GALEX band, the Sloan Digital Sky Surveys u, g, r, i, and z bands \citep{Almeida2023}, and the UKIRT J, UKIRT H, and UKIRT K$_s$ near-infrared density fluxes. We finally include the \textit{Spitzer} MIPS 24~$\mu$m and MIPS 70~$\mu$m bands, the \textit{Herschel} PACS 100~$\mu$m and 160~$\mu$m bands, and the \textit{Herschel} SPIRE 250~$\mu$m band.

\begin{table}
\caption{Photometry of 3C~326~N. The photometric bands derived from the Sloan Digital Sky Surveys (SDSS) were obtained using a circular aperture with a diameter of 8 arcsec on the SDSS DR18 \citep{Almeida2023}.}              
\label{table:phot}      
\centering                                      
\scalebox{0.95}{
\begin{tabular}{c c }          
\hline\hline                        
\textbf{Bands} & \textbf{Flux (mJy)} \\    
\hline                                   
    NUV GALEX$^\star$                     & $0.011\pm0.002$ $^{\rm a}$           \\
    \textbf{SDSS u}                         & $0.081\pm0.017$ $^{\rm b}$         \\
    \textbf{SDSS r}                         & $0.430\pm0.040$ $^{\rm b}$         \\
    \textbf{SDSS g}                         & $1.082\pm0.063$ $^{\rm b}$         \\
    \textbf{SDSS i}                         & $1.749\pm0.080$ $^{\rm b}$         \\
    \textbf{SDSS z}                         & $2.371\pm0.093$ $^{\rm b}$        \\
    UKIRT J                                 & $2.36\pm0.06$ $^{\rm c}$	         \\
    UKIRT H                                 & $2.66\pm0.07$ $^{\rm c}$	         \\
    UKIRT K$_{\rm s}$                       & $2.39\pm0.06$ $^{\rm c}$	         \\
    \textit{Spitzer} MIPS 24 $\mu$m         & $0.52\pm0.07$ $^{\rm d}$	         \\
    \textit{Spitzer} MIPS 70 $\mu$m         & $6.12\pm0.65$ $^{\rm d}$	         \\
    \textit{Herschel} PACS 100 $\mu$m       & $16.7\pm0.6$ $^{\rm d}$	         \\
    \textit{Herschel} PACS 160 $\mu$m       & $15.9\pm0.6$ $^{\rm d}$	         \\
    \textit{Herschel} SPIRE 250 $\mu$m      & $9.0\pm3.5$ $^{\rm d}$             \\
\hline                                             
\end{tabular}
}
{\small \\
{\raggedright \textbf{Notes.} \par}
{\raggedright $^\star$ \cite{Ogle10} report a NUV AB magnitude of $21.47$ (i.e., surface brightness of $\rm 1.59\times10^{-6}~W~m^{-2}~sr^{-1}$) assuming a diameter area of $5$ arcsec. The value is converted into flux density with an uncertainty equal to $20$\% of the observation. \par}
{\raggedright \textbf{References} ${\rm (a)}$~\cite{Ogle10}; ${\rm (b)}$~This work; ${\rm (c)}$~\cite{Lilly84}; ${\rm (d)}$~\cite{Guillard15}   \par}

}
\end{table}


\subsection{SED Fitting}\label{sec:cigale_code}

\begin{table*}
\caption{CIGALE modules and input parameters}
\label{table:cigale_app}      
\centering 
\scalebox{0.72}{
\begin{tabular}{c c c}     
\hline\hline       
\textbf{Parameter} & \textbf{Symbol} & \textbf{Value}\\ 
\hline                    
 \textbf{Delayed Star Formation History}   &  &  \\
e-folding time of main stellar population~(Myr) &  $\tau_\mathrm{main}$ & 250, 500, 1000, 2000, 4000, 6000, 8000,10000,12000,15000 \\
age of main stellar population~(Myr)    & age$_\mathrm{main}$ & 250, 500, 1000, 2000, 4000, 8000, 10000, 12000 \\  
\hline
  \textbf{Stellar Populations}; \cite{BC03}  & &  \\ 
 initial mass function   & IMF & \cite{Chabrier2003} \\  
metallicity    & Z & 0.008 \\ \hline
 \textbf{Dust attenuation}. Values adopted by \cite{CF00}  & &  \\   
ISM attenuation in the V-band~(mag)   & A$_\mathrm{V}^\mathrm{ISM}$ & 0.05, 0.1, 0.2, 0.3, 0.4, 0.5, 0.6 \\  
A$^{\mathrm{ISM}}_\mathrm{V}$/(A$^{\mathrm{ISM}}_\mathrm{V}+$A$^{\mathrm{BC}}_\mathrm{V}$)    & $\mu$ &  0.44\\
\hline   
  \textbf{Dust Emission}; \cite{dl2014}  &  &  \\
mass fraction of PAH    & qPAH & 1.12, 2.5, 3.19, 4.58, 5.95  \\   
minimum radiation field    & U$_\mathrm{min}$ & 5.0, 10.0, 25.0 \\  
power-law slope $\mathrm{dU/dM}\propto\mathrm{U}^\alpha$    & $\alpha$ & 2.0 \\
  fraction illuminated from $\mathrm{U_{min}}$ to $\mathrm{U_{max}}$  & $\gamma$ & 0, 0.02 \\  
\hline
 \textbf{AGN (UV-to-IR) SKIRTOR}; \cite{Stalevski2016}  &  &  \\
AGN contribution to IR luminosity    & f$_{\rm AGN}$ & 0.05, 0.1, 0.2, 0.3  \\   
Viewing angle (deg)    & $\theta$ & 60, 70, 80, 90 \\  
Polar-dust color excess (mag) & $\rm E(B-V)_{PD}$  & 0, 0.2, 0.4 \\
\hline 
\end{tabular}
}
\end{table*}

The multi-wavelength SED of 3C~326~N covering the UV-to-FIR range is fitted using the Code Investigating GALaxy Emission \citep[CIGALE\footnote{Code is available here: \url{https://cigale.lam.fr/}};][]{Boquien19, Yang20, Yang22}. CIGALE is based on an energy balance principle where ultraviolet-optical emission is partially absorbed by the dust and subsequently re-emitted in the infrared. The code is based on a Bayesian statistical analysis approach that compares millions of models of the panchromatic emission of galaxies with the observations and provides estimates of the mean values and the standard deviations of the physical parameters of the galaxy.

\begin{figure}
    \centering
     \includegraphics[width=0.49\textwidth,clip]{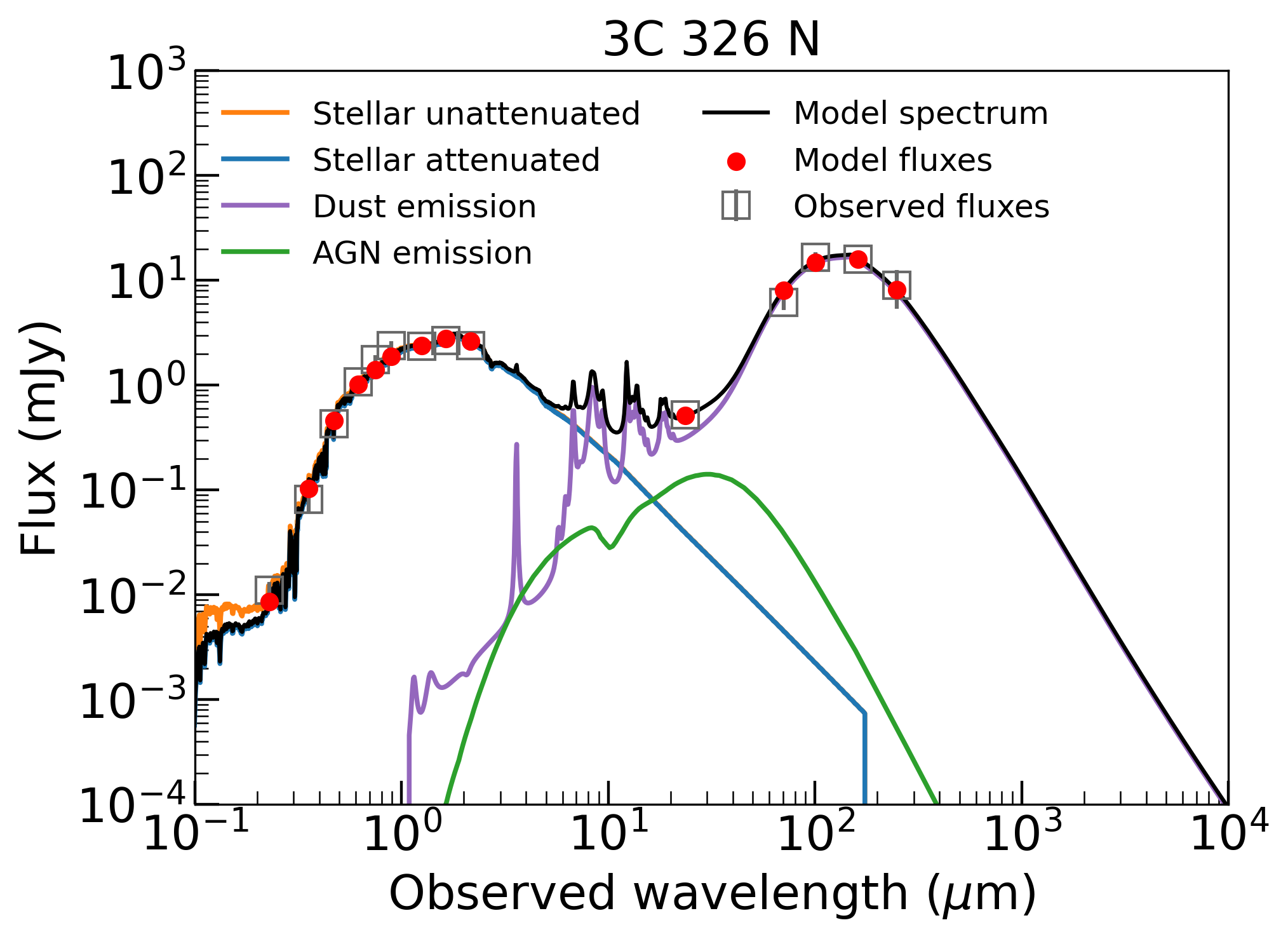}
    \caption{Spectral energy distribution of 3C~326~N. The MIR-to-FIR photometry is fitted using CIGALE. The stellar attenuated and unattenuated contributions are shown in blue and orange, respectively. The reprocessed IR emission is presented in purple and the AGN emission is in green. The best model is shown as a solid black line. The observed uncertainties are added in quadrature with a 10\% floor value to account for possible instrumental errors.}
    \label{figA1}
\end{figure}

A delayed star formation history without a burst is adopted to model this galaxy based on its low star-formation rate with the simple stellar population models of \cite{BC03} and the initial mass function of \cite{Chabrier2003}. We adopt the attenuation law recipe of \cite{CF00} and the dust IR spectral templates from \cite{dl2014}. In contrast with the results presented in \cite{Lanz2016}, we include the IR contribution of the AGN using the SKIRTOR clumpy torus models of \cite{Stalevski2016} and following the configuration reported in \cite{Yang22} with a fixed opening angle of 40$^\circ$ and varying type-2 viewing angles ($\theta$). All the parameters used for the fit are summarized in Table \ref{table:cigale_app}.

\subsection{Physical parameters: 3C~326~N}\label{sec:cigale_code_results}

Fig. \ref{figA1} shows the results of the fit of the SED. The black solid line corresponds to the best SED and the different color lines show the contribution of the stellar, dust, and AGN emission to the overall SED of the galaxy. We obtain a reduced $\chi^2$ of $1.2$. The Bayesian estimated values of the stellar mass and SFR are presented in Table \ref{table_bayes_app} with new estimations of the UV and IR luminosities integrated between $0.0912$ and $0.24~\mu{\rm m}$ and between $1$ and $1000~\mu{\rm m}$, respectively. The visual attenuation, $\rm A_v$, and the relative contribution of the AGN to the IR emission, $\rm f_{AGN}$, are also reported. The derived stellar mass and SFR are in good agreement (within 30\%) with the values reported in the literature \citep{Ogle10, Nesvadba10, Guillard15}. The IR luminosity is also consistent with the fits performed by \cite{Guillard15}.

\begin{table}
\caption{Physical parameters derived from the fit of the SED.}              
\label{table_bayes_app}      
\centering                                      
\begin{tabular}{l c }          
\hline\hline                        
\textbf{Parameter} & \textbf{Value} \\    
\hline                                   
    $\rm SFR~(M_\odot~yr^{-1})$  & $(5\pm3)\times10^{-2}$ \\
    $\rm M_{*}~(M_\odot)$        & $(0.7\pm0.1)\times10^{11}$ \\
    $\rm L_{IR}~(L_\odot)$       & $ (7.1\pm1.4)\times10^9$    \\
    $\rm L_{UV}~(L_\odot)$       & $ (8.5\pm0.4)\times10^8$    \\
    $\rm A_{V}~(mag)$            & $0.1$                        \\
    $\rm f_{AGN}$                & $0.1$                       \\
    
\hline                                             
\end{tabular}
\end{table}

\section{Alternative scenarios}\label{ssec:OnlyShockPDR}

\subsection{PDR-only scenario}\label{ssec:fitting_PDRS}

\begin{figure*}[!ht]
    \centering
     \includegraphics[width=0.99\textwidth,clip]{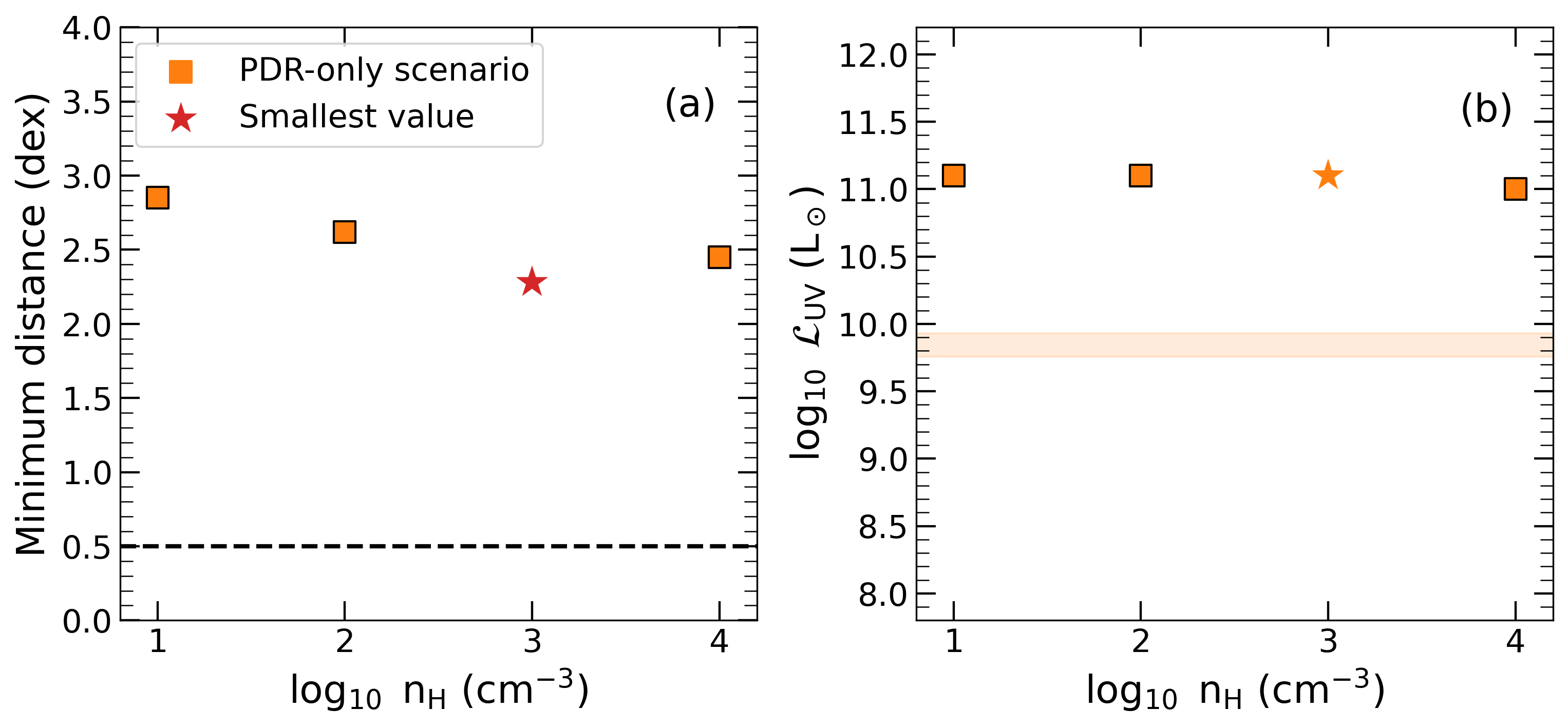}\\
     \includegraphics[width=0.99\textwidth,clip]{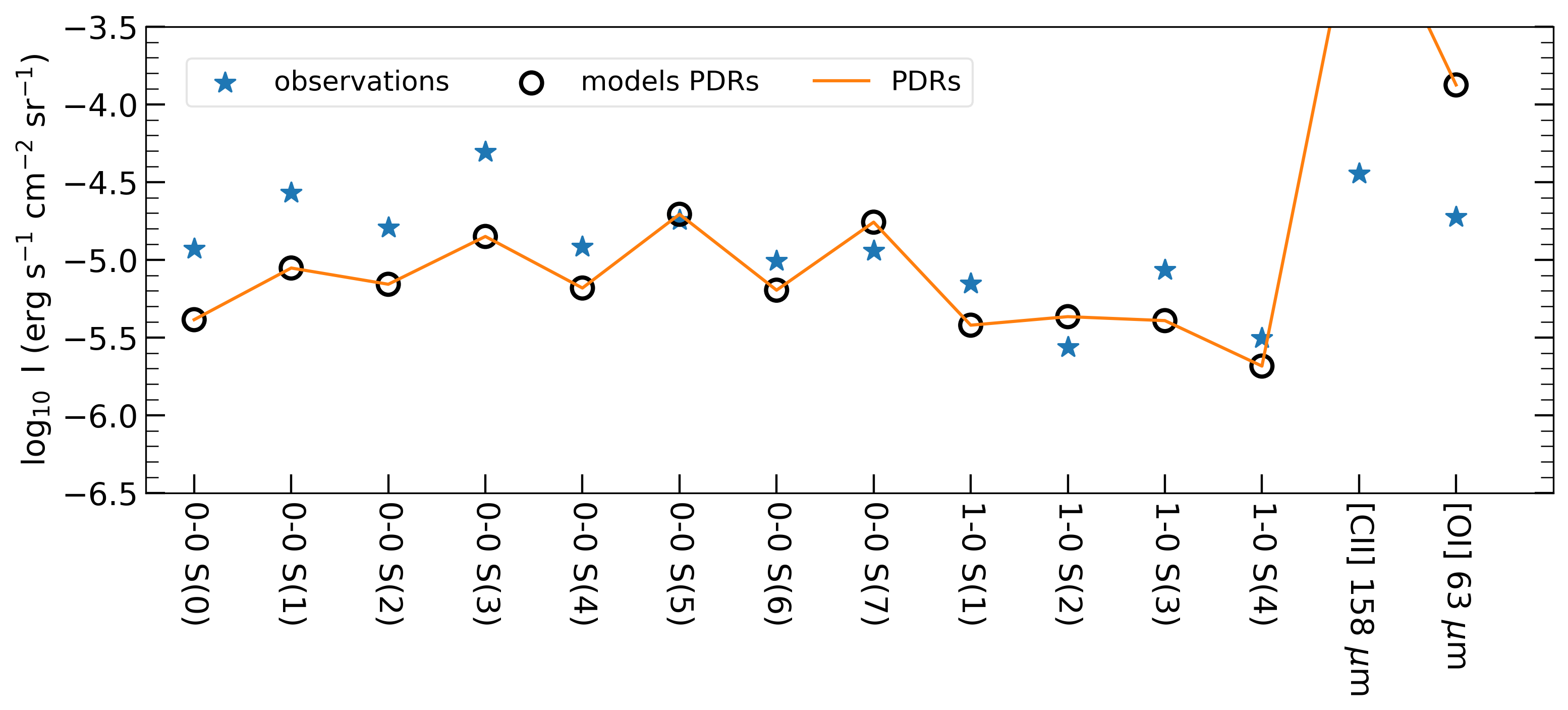}
    \caption{PDR only scenario: summary of the minimum distance, the UV-reprocessed luminosity, and the intensities predicted with a sole distribution of PDR models. \textbf{Panel (a)}: minimum distance as a function of the gas density. The dash-line represents a 0.5 dex limit (i.e., a factor of three difference between observed and modeled line intensities) to guide the eye. \textbf{Panel (b)}: UV-reprocessed luminosity as a function of the gas density. The orange shaded area represents the SED estimation of the reprocessed UV-luminosity (see Appendix \ref{ssec:SED}). \textbf{Panel (c)}: comparison between the observed line intensities and those obtained for the model at $\rm n_H = 10^3~cm^{-3}$, $\rm b = 0.1$, $\rm G_0 = 10$. Observations are presented as blue stars (see Table \ref{table:1}) and modeled intensities are shown as open black circles. In the lower panel, the residuals are shown as orange squares. The dash and solid lines give the $\rm \pm 0.5, 1~dex$ limits, respectively.}
    \label{fig_pdronly}
\end{figure*}

We test the possibility of reproducing the ensemble of observations in 3C~326~N using only PDR models. The impact of the PDR density on the interpretation of observations is shown on panel (a) of Fig. \ref{fig_pdronly} which displays the minimum distance between the observations and the distribution of PDR models for a density ranging from $10$ to $\rm 10^4~cm^{-3}$. The panel (b) shows the UV-reprocessed luminosity as a function of the density and the panel (c) displays the comparisons between the observations and the intensities of the lines predicted by the model at $\rm n_H = 10^3~cm^{-3}$ (which corresponds to the smallest distance). Fig.~\ref{fig_pdronly} shows that sole PDR models cannot explain the observational dataset and raise two main issues. First, the minimum value of the distance is large ($\sim 2.3$). As shown in panel (c), the need to reproduce the observed intensities of all the rovibrational lines of H$_2$ leads to overestimations of the intensities of the [CII]~$158~\mu$m and [OI]~$63~\mu$m lines by one to two orders of magnitude. Second, the predicted UV-reprocessed luminosity is found to be more than one order of magnitude larger than that deduced from the IR luminosity of the galaxy (see panel b). This result demonstrates the importance of estimating the radiative energy budget of galaxies for the interpretation of atomic and molecular emission lines.

\subsection{Shock-only scenario}\label{ssec:fitting_shock}

\begin{figure*}[!ht]
    \centering
     \includegraphics[width=0.99\textwidth,clip]{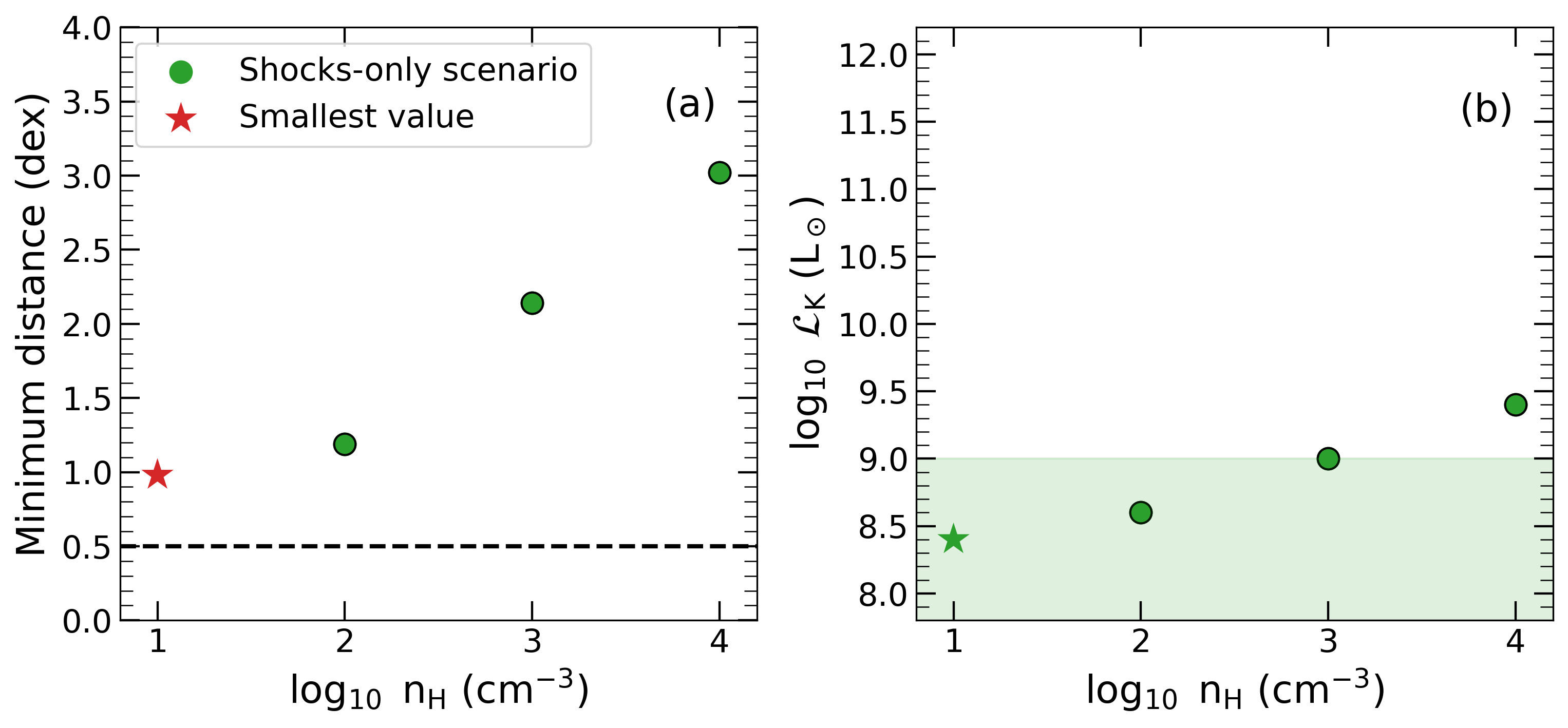}\\
     \includegraphics[width=0.99\textwidth,clip]{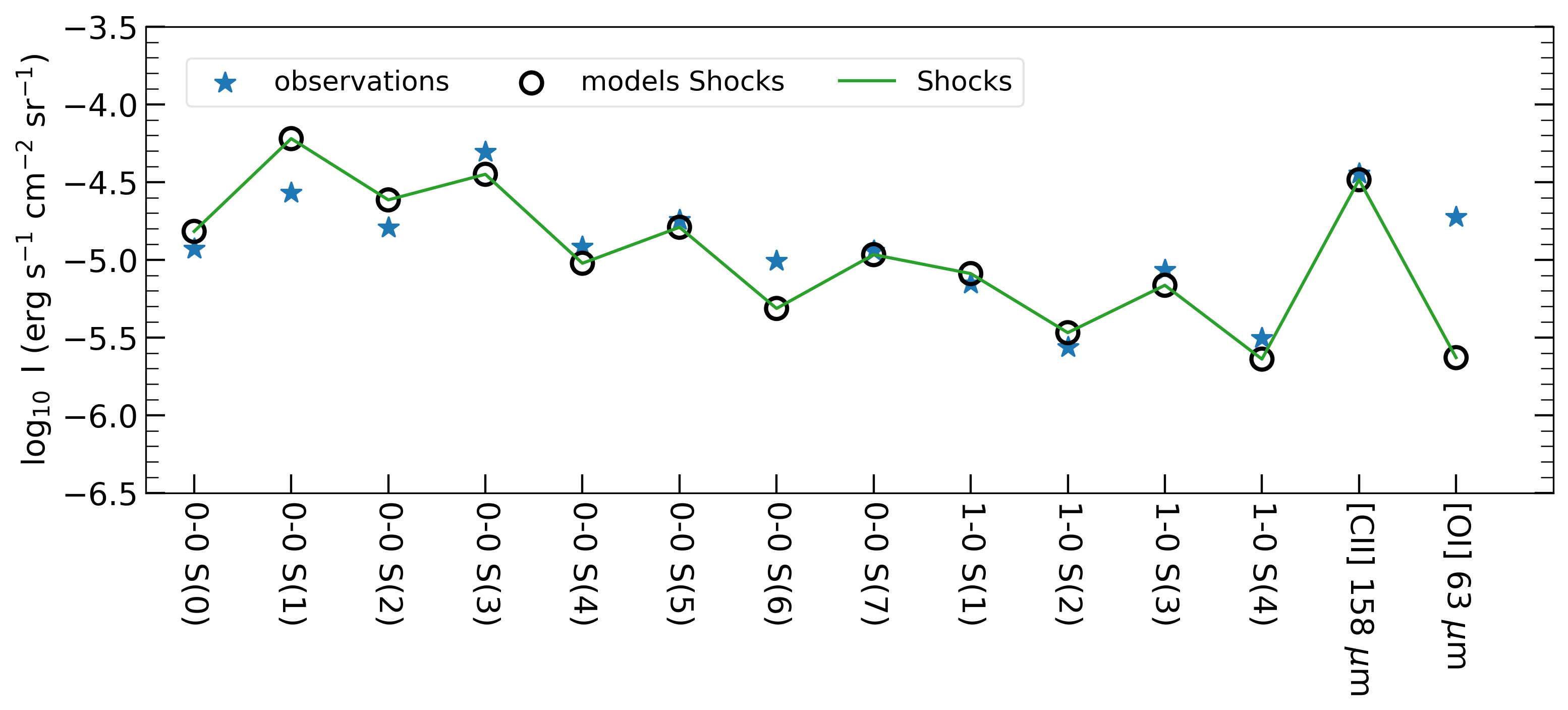}
    \caption{Shock only scenario: summary of the minimum distance, the mechanical luminosity, and the intensities predicted with a sole distribution of shock models. \textbf{Panel~(a)}: minimum distance as a function of the pre-shock density. The dash-line represents a 0.5 dex limit (i.e., a factor of three difference between observed and modeled line intensities) to guide the eye. \textbf{Panel~(b)}: mechanical luminosity as a function of the pre-shock density. The green area represents $<1$\% of the estimated jet kinetic luminosity of 3C~326~N (see Sect.~\ref{ssec:result_energy}). \textbf{Panel~(c)}: comparison between the observed line intensities and those obtained for the model at $\rm n_H = 10~cm^{-3}$, $\rm b = 0.1$, $\rm G_0 = 10$. Observations are presented as blue stars (see Table \ref{table:1}) and modeled intensities are shown as open black circles. In the lower panel, the residuals are shown as green squares. The dash and solid lines give the $\rm \pm 0.5, 1~dex$ limits, respectively.}
    \label{fig_shockonly}
\end{figure*}

We also consider the possibility of reproducing all the observations using only shock models. The impact of the density of the pre-shock gas on the interpretation of observations is shown in Fig. \ref{fig_shockonly} (panel a) which displays the minimum distance between the observations and the distribution of shocks models for a density ranging from $10$ to $\rm 10^4~cm^{-3}$. The panel~(b) shows the mechanical luminosity as a function of the density and the panel~(c) displays the comparisons between the observations and the intensities of the lines predicted by the model at $\rm n_H = 10~cm^{-3}$ (which corresponds to the smallest distance). Interestingly, the predicted mechanical luminosity is found to be in agreement with the available energy deduced from the jet's power as long as the density is below $\rm 10^3~cm^{-3}$ (see panel b). Fig.~\ref{fig_pdronly} shows, however, that a sole distribution of shocks cannot explain the entire observational dataset. As shown in panel (c), the need to reproduce the observed intensities of the rovibrational lines of H$_2$ and of the fine structure line of C$^+$ leads to an underestimation of the [OI] emission by about one order of magnitude. This result demonstrates the importance of including both atomic and molecular lines for the interpretation of extragalactic observations.


\subsection{Gaussian distribution of the shock velocities}\label{ssec:GaussianSol}

\begin{figure*}[!ht]
    \centering
     \includegraphics[width=0.99\textwidth,clip]{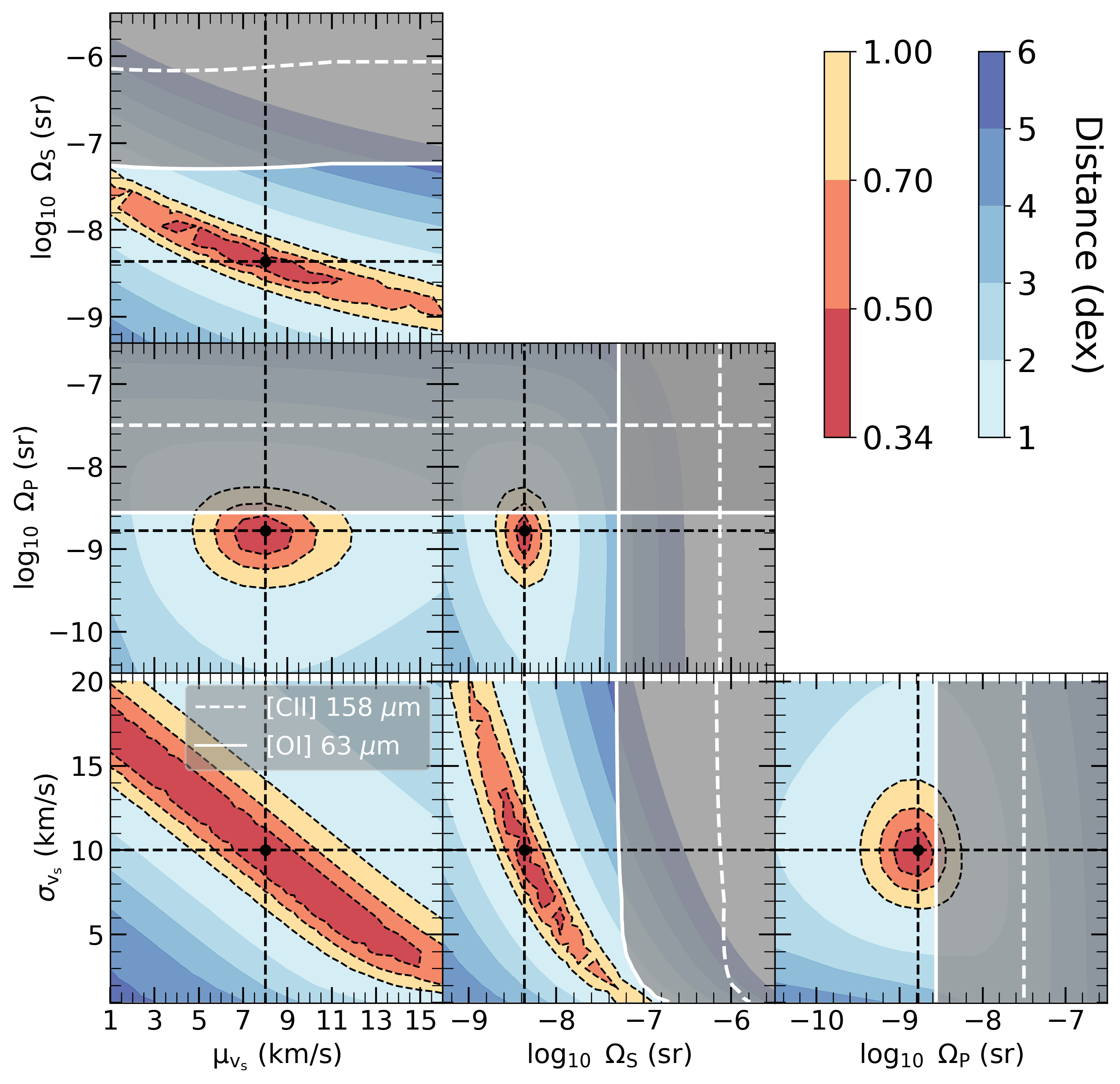}
    \caption{2D cutoff of the likelihood distributions of the parameters describing the shock and the PDR distributions obtained for a Gaussian distribution of shock velocities. The parameters are the total shock solid angle $\Omega_{\rm S}$, the central shock velocity $\mu_{\rm v_s}$ and the dispersion of the shock velocities $\sigma_{\rm v_s}$, and the PDR total solid angle $\Omega_{\rm P}$. The color code represents the distance in dex units. The $0.5$, $0.7$, and $1.0$ dex limits are highlighted with dashed black lines. The straight black dashed lines and the black points indicate the position of the global minimum. Gray areas represent the regions where the assumption on the radiative transfer regarding cross absorption between surfaces breaks down. Note that the criterion is more stringent for the [OI]~$63~\mu$m line (gray area above the white solid line) than for the [CII]~$158~\mu$m line (gray area above the white dashed line) and is always fulfilled for H$_2$ lines (see Appendix~\ref{sec:Lehman_toymodel}).} 
    \label{fig0_2_app}
\end{figure*}

We finally explore the impact of the form of the probability distribution function $f_{\rm S}$ assuming that the shock velocities follow a Gaussian distribution
\begin{equation}
f_{\rm S} = \frac{1}{\sqrt{2\pi} \sigma_{\rm v_s}} {\rm exp} \left( - \frac{(\rm v_s - \mu_{\rm v_s})^2}{2 \sigma_{\rm v_s}^2}\right)
\end{equation}
rather than an exponential distribution. Fig.~\ref{fig0_2_app} shows the distance computed with Eqs.~\ref{eq:4} and \ref{eq:4_distance} as a function of the parameters of the Gaussian distribution of shocks, $\rm \Omega_S$, $\mu_{\rm v_s}$, and $\sigma_{\rm v_s}$, and the parameter of the Dirac distribution of PDRs, $\Omega_{\rm P}$. 

The most striking feature of Fig.~\ref{fig0_2_app} is the strong degeneracy between $\Omega_S$, $\mu_{\rm v_s}$, and $\sigma_{\rm v_s}$. A narrow distribution of shock velocities ($\sigma_{\rm v_s} \sim 6$~km~s$^{-1}$) centered at $\mu_{\rm v_s} = 12$~km~s$^{-1}$ leads to the same distance than a broad distribution ($\sigma_{\rm v_s} \sim 15$~km~s$^{-1}$) centered at $\mu_{\rm v_s} = 3$~km~s$^{-1}$. This strong degeneracy is not surprising and is simply due to the fact that the intensities of the rovibrational lines of H$_2$ are mostly emitted by shocks with velocities between 6 and 15~km~s$^{-1}$, regardless of the number of shocks below 6~km~s$^{-1}$ (see Fig.~\ref{fig4_1}).

\begin{figure*}
    \centering
     \includegraphics[width=0.99\textwidth,clip]{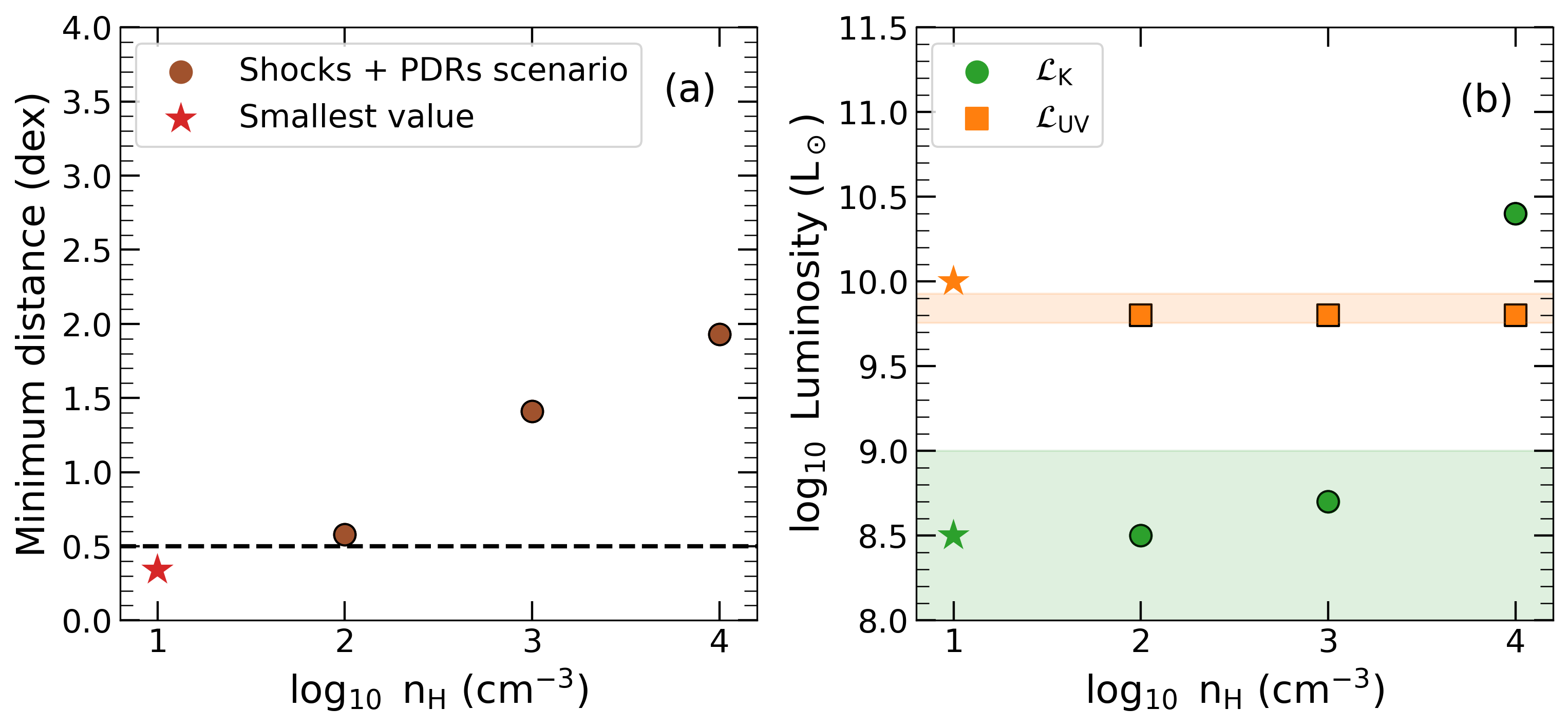}\\
     \includegraphics[width=0.99\textwidth,clip]{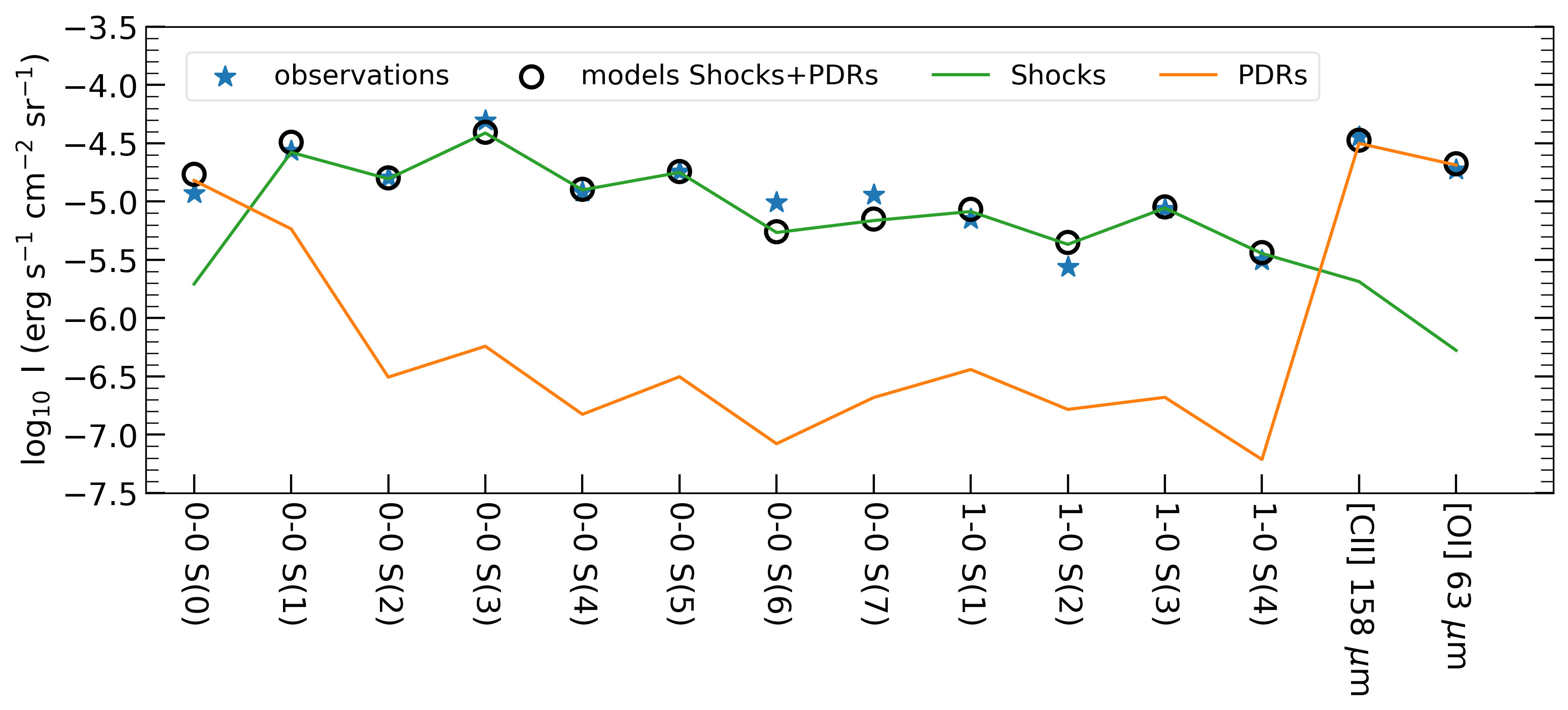}
    \caption{
    PDR and shock scenario with a Gaussian distribution of shock velocities: summary of the minimum distance, the UV-reprocessed and mechanical luminosities, and the intensities predicted. \textbf{Panel~(a)}: minimum distance as a function of the gas density. The dash-line represents a 0.5 dex limit (i.e., a factor of three difference between observed and modeled line intensities) to guide the eye. \textbf{Panel~(b)}: UV-reprocessed luminosity (orange) and mechanical luminosity (green) as functions of the gas density. The orange shaded area represents the SED estimation of the reprocessed UV-luminosity (see Appendix \ref{ssec:SED}). The green area represents $<1$\% of the estimated jet kinetic luminosity of 3C~326~N (see Sect.~\ref{ssec:result_energy}). \textbf{Panel~(c)}: comparison between the observed line intensities and those obtained for the model at $\rm n_H = 10~cm^{-3}$, $\rm b = 0.1$, $\rm G_0 = 10$. Observations are presented as blue stars (see Table \ref{table:1}) and modeled intensities are shown as open black circles. In the lower panel, the residuals are shown as green squares. The dash and solid lines give the $\rm \pm 0.5, 1~dex$ limits, respectively.}
    \label{shockpdr_app}
\end{figure*}

The impact of the gas density on the interpretation of observations is shown in Fig. \ref{shockpdr_app} (panel a) which displays the minimum distance between the observations and the distribution of shocks and PDRs for a density ranging from $10$ to $\rm 10^4~cm^{-3}$. The panel~(b) shows the UV-reprocessed luminosity and the mechanical luminosity as functions of the density and the panel~(c) displays the comparisons between the observations and the intensities of the lines predicted by the model at $\rm n_H = 10~cm^{-3}$ (which corresponds to the smallest distance).

This ensemble of figures is remarkably similar to those obtained for an exponential distribution of shock velocities (see Fig. \ref{figshockandPDR}). In particular, they indicate that the emission likely originates from the diffuse interstellar gas ($\rm n_H \leqslant 100$ cm$^{-3}$) and that the 0-0 S(0) line of H$_2$ and the fine structure lines of C$^+$ and O are mostly emitted by PDRs while all the other H$_2$ lines are emitted by shocks, with radiative and mechanical energy budgets in agreement with observational constraints. All these results confirm the interpretation obtained in the main text with an exponential distribution of shock velocities. Fig. ~\ref{fig0_2_app} reveals that assuming a more complex distribution only leads to additional degeneracies. This means that the current ensemble of observations is insufficient to discriminate different distributions of shock velocities.


\end{appendix}

\end{document}